\documentclass[aps,prl,twocolumn,superscriptaddress,10pt]{revtex4-2}
\usepackage{amsthm}
\usepackage{ulem,mathptmx}
\usepackage[pdftex,dvipsnames,usenames]{xcolor}
\usepackage{tikz}
\usepackage{bbm, dsfont}
\usetikzlibrary{calc}
\usetikzlibrary{shapes.geometric, arrows}
\usepackage{eufrak}

\usepackage{enumitem}
\usepackage{mathtools}
\usepackage{graphicx}

\usetikzlibrary{positioning}
\usepackage{xr-hyper}
\usepackage[colorlinks=true,citecolor=myblue,linkcolor=myblue,urlcolor=myblue]{hyperref}
\newcommand{\bematrix}{\left(\begin{matrix}}
\newcommand{\ematrix}{\end{matrix}\right)}

\usepackage{ulem} 
\usepackage{amsmath,amssymb}
\usepackage{bm}
\usepackage{bbold} 
\usepackage{color}
\usepackage[english]{babel}
\usepackage{graphicx}

\usepackage{amsmath}
\usepackage{amssymb}
\usepackage{amsfonts}
\usepackage{natbib}
\usepackage{dsfont}

\usepackage{float}

\definecolor{myblue}{rgb}{0.2,0.2,0.8}

\newenvironment{proof-of}[1]{\medskip\noindent\textbf{Proof of {#1}.}}{\hfill$\blacksquare$\medskip}
\newcommand{\ket}[1]{\left\vert#1\right\rangle}
\newcommand{\bra}[1]{\left\langle#1\right\vert}

\usepackage{colortbl}
\usepackage{xcolor}
 
\usepackage{tcolorbox}

\makeatletter
\DeclareRobustCommand{\mathbbm}[1]{%
  \ifnum1=\pdfstrcmp{#1}{1}
    \mathrm{\usefont{U}{bbold}{m}{n}1}
  \else
    \ifnum0=\pdfstrcmp{#1}{0}
      \mathrm{\usefont{U}{bbold}{m}{n}0}
    \else
      \@latex@error{Comando \mathbbm non supportato per #1}\@ehd
    \fi
  \fi
}
\makeatother

\begin{document}

\title{Unified boson sampling}

\author{Luca Bianchi}
\affiliation{Department of Physics and Astronomy, University of Florence, 50019, Firenze, Italy}
\email{luca.bianchi@unifi.it}

\author{Carlo Marconi}
\affiliation{Istituto Nazionale di Ottica del Consiglio Nazionale delle Ricerche (CNR-INO), 50125 Firenze, Italy}

\author{Laura Ares}
\affiliation{Theoretical Quantum Science, Institute for Photonic Quantum Systems (PhoQS), Paderborn University, Warburger Stra\ss{}e 100, 33098 Paderborn, Germany}

\author{Davide Bacco}
\affiliation{Department of Physics and Astronomy, University of Florence, 50019, Firenze, Italy}

\author{Jan Sperling}
\affiliation{Theoretical Quantum Science, Institute for Photonic Quantum Systems (PhoQS), Paderborn University, Warburger Stra\ss{}e 100, 33098 Paderborn, Germany}

\begin{abstract}
    Boson sampling is a key candidate for demonstrating quantum advantage, and has already yielded significant advances in quantum simulation, machine learning, and graph theory.
    In this work, a unification and extension of distinct forms of boson sampling is developed.
    The devised protocol merges discrete-variable scattershot boson sampling with continuous-variable Gaussian boson sampling.
    Thereby, it is rendered possible to harness the complexity of more interesting states, such as squeezed photons, in advanced sampling protocols.
    A generating function formalism is developed for the joint description of multiphoton and multimode light undergoing Gaussian transformations.
    The resulting analytical tools enable one to explore interfaces of different photonic quantum-information-processing platforms.
    A numerical simulation of unified sampling is carried out, benchmarking its performance, complexity, and scalability.
    Entanglement is characterized to exemplify the generation of quantum correlations from the nonlinear interactions of a unified sampler.
\end{abstract}

\date{\today}

\maketitle    

    \noindent
    \paragraph{Introduction.---}
        The development of quantum information processing \cite{nielsen2001quantum} has been driven by the pursuit of physical platforms capable of efficiently encoding, manipulating, and transmitting quantum states while maintaining coherence and scalability \cite{bennett2000quantum,khan}. Among the various candidate systems, quantum optical/photonic platforms \cite{mandel1995optical, vogel2006quantum} offer unique advantages due to the weak interaction of photons with their environment \cite{photoniccompreview,flamini2018photonic}. This intrinsic robustness against decoherence, coupled with well-established optical control techniques, has lifted photonics as a key architecture for quantum communication, quantum computing, quantum sensing, and quantum simulation \cite{silverstone2016silicon, obrien2009photonic}.
        
        Aside from single photons, which are the building blocks for discrete-variable optical quantum information \cite{kok2007linear}, a particularly important class of states is squeezed states, which emerge from the family of Gaussian transformations that redistribute quantum uncertainty between conjugate quadratures of the optical field \cite{schnabel2017squeezed}. Such Gaussian states play a key role in continuous-variable quantum information \cite{lloyd,brauvanloock,weedbrook2012Gaussian}. The Gaussian nature of these states allows for deterministic state preparation and transformation via linear optics and homodyne detection \cite{zhang2018integrated}.
        
        Among all quantum technologies, boson sampling, introduced by Aaronson and Arkhipov \cite{aaronson2011computational}, is a quantum optical implementation of a computational model that may surpass classical algorithms, overcoming the extended Church-Turing Thesis \cite{churchturing} and reaching quantum advantage.        In its original version, the protocol consists of sending single-photon states through a randomly selected linear interferometer and performing measurements at the output. The sampled probability distribution, proportional to the modulus squared permanent of the scattering matrix, falls in the $\#\mathrm{P}$ complexity class \cite{permanentcomplexity}. However, boson sampling experiments mostly rely on probabilistic photon sources \cite{tillmann2013experimental}, limiting the boson sampling scalability. To address this issue, many alternative models were introduced: Scattershot boson sampling (SBS) \cite{bentivegna2015experimental,scattershot} improves the efficiency by using heralded photon pairs, while Gaussian boson sampling (GBS) \cite{Hamilton2017Gaussian, detailedGaussianHamilton, zhong2019experimental} leverages squeezed states for the production of a large number of indistinguishable photons. In the GBS, the Hafnian of the scattering matrix \cite{combinatorialpmp,rudelson2016hafnians} built by the measured  modes provides the required $\#\mathrm{P}$ complexity. Further applications of GBS can be found in molecular vibronic spectra simulation \cite{huh2015boson}, graph-based problems \cite{graph1, graph2}, state preparation \cite{sabapathy2019production, bourassa2021blueprint} and machine learning tasks \cite{machine_learning1, machine_leanring2}.
        
        More recently, however, many quantum information protocols were designed on hybrid platforms where both discrete and continuous properties are exploited \cite{andersen2015hybrid}. Integrating squeezing transformations into large-scale photonic quantum networks presents an opportunity to take advantage of the best of both encodings by entering the domain of non-Gaussian quantum states \cite{walschaers2021non}.
        
        In this work, we introduce a hybrid protocol for boson sampling. The protocol entails feeding single photons to optical parametric amplifiers, followed by a Haar random linear interferometer and photodetection. To describe this unified boson sampler, we firstly develop an exact formalism comprising both Gaussian and photonic resources that harnesses the expectation values for momenta to arbitrary order. Our method represents photons via derivatives of a Gaussian generating function that carries all the information about the time evolution of the multiphoton system. We prove that the unified boson sampler encompasses early sampling models, and we validate this statement by determining the distance from SBS and GBS. Furthermore, we benchmark the complexity of the here-devised sampler by studying the computational time as a function of the system size. We finally compute the entanglement created across the modes as a function of the mean total photon number, highlighting the dependency on the input photons as well as on the squeezing intensity.\\ 
        
    \paragraph{Methods.---}
 
        In order to describe the dynamics of an $M$-mode optical field, we introduce the vector of canonical variables in phase space \cite{serafini2023quantum}, $\boldsymbol{\hat{a}}^{\mathrm{T}}= [\hat{a}_1,\dots,\hat{a}_{M},\hat{a}^{\dagger}_1,\dots,\hat{a}^{\dagger}_{M}]$, in such a way that canonical commutation relations can be recast as $
            [\boldsymbol{\hat{a}}, \boldsymbol{\hat{a}}^{\mathrm{T}}] = i\boldsymbol{J} $
        where
        $
            \boldsymbol{J}
            =
            -i\begin{bmatrix}
                    \mathbbm{1} & \mathbb{0} \\
                    \mathbb{0} & -\mathbbm{1}
            \end{bmatrix},
        $ with $\mathbb{0}$ and $\mathbbm{1}$ respectively denoting the null matrix and the identity. (Note that $\hat{I}$ is the identity operator acting on quantum states.)
        Throughout the work, bold quantities denote $2M$-dimensional objects in the space of canonical variables.
        
        All Gaussian transformations preserve canonical commutation relations \cite{Gaussianqinfo}, are closed under matrix multiplication, and form the complex symplectic group \cite{sympgroup, bookliegroups} 
        $
            \mathrm{Sp}_{\mathbb{C}}(2M) 
            =
            \{ 
                \boldsymbol{T} \in \mathbb{C}^{2M \times 2M} | \; 
                \boldsymbol{TJT}^{* \mathrm{T}} 
                =
                \boldsymbol{J} 
            \},
        $
        where we denote with $* \mathrm{T}$ the Hermitian conjugation for phase-space matrices, while reserving the $\dagger$ symbol for operators acting on the Hilbert space of states.
        Therefore, any unitary operator $\hat{T}$ as a Gaussian transformation in Hilbert space can be expressed by the action of a complex symplectic matrix on vectors of canonical variables
        $
            \hat{T}\boldsymbol{\hat a} \hat{T}^{\dagger} 
            =
            \boldsymbol{T}\boldsymbol{\hat a}.
        $
        Further, a generic $\boldsymbol{T} \in \mathrm{Sp}_{\mathbb{C}}(2M)$ can always be written as four blocks made by $M \times M$ complex matrices \cite{Gaussianops}
        $
            \boldsymbol{T} 
            =
            \begin{bmatrix}
                U & V \\
                V^\ast & U^\ast
            \end{bmatrix}
        $
        subject to the symplectic constraints
        $
            U U^{* \mathrm{T}}-V V^{* \mathrm{T}} 
            = \mathbbm{1} \, , \; \; \; \;
            U V^{\mathrm{T}}= V U^{\mathrm{T}}.
        $
        Linear interferometry ($V=\mathbb{0}$) is now identified with the group of unitary transformations $\mathrm{U}(M)$, which turns out also to be isomorphic to the maximal compact subgroup of $\mathrm{Sp}_{\mathbb{C}}(2M)$ \cite{matrixdecompositions}.\\ 
        We may decompose Gaussian matrices with the so called Bloch-Messiah decomposition \cite{simonbmdecomposition} as
        $
           \boldsymbol{T}=\boldsymbol{LS R}
        $,
        with $\boldsymbol{L},\boldsymbol{R}$ being $M$-modes unitaries and $\boldsymbol{S}$ taking the form
        $
            \boldsymbol{S} 
            = 
            \begin{bmatrix}
                \sigma_{U} &\sigma_{V}\\
                \sigma_{V}^\ast & \sigma_{U}^\ast
            \end{bmatrix},
        $
        with $\sigma_U = \mathrm{diag}(\sigma_{u_1},\dots,\sigma_{u_{M}})$ and $\sigma_V=\mathrm{diag}(\sigma_{v_1},\dots,\sigma_{v_{M}})$ containing the $2M$ singular values of $\boldsymbol{T}$, obeying $\sigma_{u_i}=\sqrt{1+|\sigma_{v_i}|^2}$. This identity stems from the singular value decomposition \cite{horn2012matrix} when applied to symplectic matrices.
        Thus, Bloch-Messiah states how any Gaussian operation can be decomposed into a linear interferometric transformation (unitary rotation), followed by single-mode squeezing (diagonal matrices) and another linear transformation \cite{braunstein2005squeezing}.
        
        To compactly describe input photons, we refer to the formalism of generating functions \cite{mcbride2012obtaining}. For $M$-modes photon states, it turns out that the matrix $\bigotimes_{k=1}^{M}\ket{n_k}\bra{n_k}$ can be equivalently described by the operator
        $
            \hat{\Pi}_{\bar{n}} 
            =
            \left.
                \frac{1}{\bar n!}
                \prod_{k=1}^{M}
                     \partial_{x_k}^{n_k} \hat{E}(x_k)
            \right|_{\bar x= \bar 0}
        $
        , with
        $
            \hat{E}(x_k) 
            = 
            x_k^{\hat{n}_k}
            =
            :e^{
                -(1-x_k)\hat{n}_k
                }:
        $,  
        and ``$:\cdots:$'' denoting the normal ordering prescription \cite{normalordering}. Furthermore, we utilize a multi-index notation, namely $\bar{x}= (x_1,\dots,x_M)$. For each mode $k$, $\hat{E}(x_k)$ simultaneously encodes the full information of all photon-number states. Such compact writing allows for efficient computations of statistical moments of the field and was proven to be also useful for the description of thermal states, Fock states and detector performances characterization \cite{sperlingdetectorchar, sperlingquantuengineering}.
        
        To achieve the normally-ordered form of matrices describing propagated input states, we rely on the group theoretical properties of Gaussian Hamiltonians.
        In particular, for $M$-modes Gaussian transformations ladder operators turns out to be an infinite dimensional representation of an $\mathfrak{sp}(2M+2)$ subalgebra \cite{dictionaryliealgebras, gilmorecoherent}. Thus, tacking operations on group elements that only involve products is representation-independent \cite{knapp1996lie,zee2016group}, meaning that the reordering of operators can be performed into a simpler, finite-dimensional representation. Combining the aforementioned technique with the Bloch-Messiah decomposition leads to a normally-ordered form for
        \begin{equation}
            \label{nodin}
                \hat{T}
                =
                e^{-\frac{1}{2}\hat{a}^{\dagger}U^{-*\mathrm{T}} V^{\mathrm{T}}\hat{a}^{\dagger}}
                e^{-\log{U^{*\mathrm{T}}}\left(\hat{a}^{\dagger}\hat{a}+\hat{I}/2\right)}
                e^{\frac{1}{2}\hat{a}V^{*\mathrm{T}}U^{-*\mathrm{T}}\hat{a}},
        \end{equation}
        see Supplemental Material (SM) \cite{sm}.
        Previously, similar techniques were shown to be useful, e.g., in Refs. \cite{schwinger2000quantum,caves2020su11,disentanglement}. 
        From the above result, we can compute the normally-ordered time evolution of photon-number states:
        \begin{equation}
            \label{measurement}
                \left.
                    \hat{\Pi'}_{\bar{n}}
                    =
                    \partial^{\bar{n}}_{\bar{x}}
                    :\frac{
                            \exp{ 
                                \left\{
                                    \frac{1}{2}\boldsymbol{\hat{a}}^{\mathrm{T}}
                                    \begin{bmatrix}
                                        A & B \\
                                        B^{\mathrm{T}} & A^\ast
                                    \end{bmatrix}
                                    \boldsymbol{\hat{a}}  
                                \right\}
                            }
                        }
                    {
                        {\bar{n}!}|\det[U]|\sqrt{
                            \prod_{i=1}^{M}
                                D_i
                        }
                }:
                \right|_{\bar x=\bar 0},
        \end{equation}
        with
        \begin{equation}
            \begin{split}
                  &D= \mathrm{diag}\left(
                           \mathbbm{1}-XU^{-*}V^{*}XV^{\mathrm{T}}U^{-\mathrm{T}}
                        \right)\\
                  &A = -V^{*}U^{-} + U^{-\mathrm{T}}XD^-V^{*\mathrm{T}}U^{-*\mathrm{T}}XU^{-}\\ 
                  &B = -\left(
                            \mathbbm{1} -U^{-*\mathrm{T}}XD^-U^{-}
                        \right) \; ,
            \end{split} 
        \end{equation}
        where $X = \mathrm{diag}\left(x_1,\dots,x_M\right)$ and a superscript ``$-$'' denoting the inverse, cf. SM \cite{sm}. Next, we show how this formalism unifies the description of any process involving generation, transformation, and detection of any photon-number state combined with Gaussian resources, thus subsuming previous results, e.g. in Refs. \cite{bianchi2025predetection,engelkemeier2020quantum}. Notice that a complementary analysis of the above scenario has been developed in terms of loop Hafnians in \cite{quesada2019franck, miatto2020fast, cardin2022photon}.  

\paragraph{Unified boson sampling.---}

        Having developed a formalism to analytically describe Gaussian transformations of photon states, we can incorporate both the SBS and the GBS in our model and describe the physics that lies in between; i.e., we can give a joint description of these boson sampling problems, a protocol that we dub unified boson sampling (UBS). For simplicity, let $N_{\mathrm{in}} = N_{\mathrm{out}} = N$ single input and output photons. By calling $ \boldsymbol{G} $ the matrix in the exponent of Eq. \eqref{measurement}, it can be shown (SM \cite{sm}) that probability outcomes are described by the following expression:
            \begin{equation}
                \label{probformulaubs}
                \begin{aligned}
                    P_{N \rightarrow N}
                    =
                    \frac{1}{|\det[U]|}
                    \sum_{S}
                    {}&
                    \mathrm{Haf}
                    \left(
                        |U^{-1} V |^{\circ 2}_{(\bar{\beta}, \bar{\beta})}
                    \right)
                    \\
                    {}&\times
                    \left.
                            \frac{\partial^{N-|\bar{\beta}|}}{\prod_{i \notin S}\partial x_i}
                            \mathrm{Haf}
                            \left(
                                \boldsymbol{G}^{S}
                            \right)
                    \right|_{\bar{x} = \bar{0}},
                \end{aligned}
    \end{equation}
        where $S$ runs over the sets $\{1,\dots,N\}$ and $|\,\cdots\,|^{\circ 2}$ stands for the elementwise modulus-square of a matrix. Given $\bar \beta$ the subset of derivation variables, the notation $(\bar{\beta},\bar{\beta})$ stems for the subtraction of the $\bar{\beta}-$rows and columns. $\boldsymbol{G}^S$ indicates the submatrix selected by the modes in which photons are detected at the output.  
        
\begin{figure}
            \centering
            \includegraphics[width=1\linewidth]{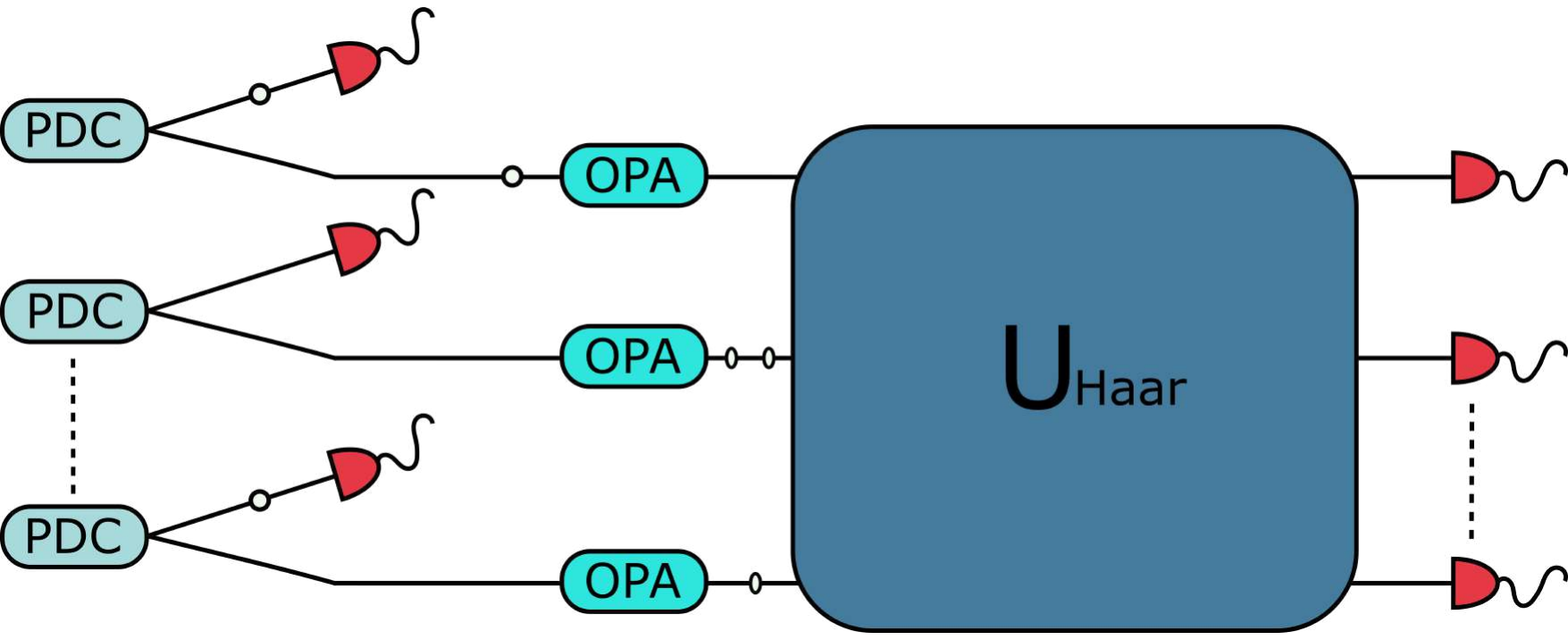}
            \caption{%
                Implementation of an unified boson sampler. Input photons, generated via parametric-down-conversion processes (PDCs) and conditional measurements, are fed into optical parametric amplifiers (OPAs) to generate squeezed photons. The latter are mixed by a Haar-random unitary ($U_{\mathrm{Haar}}$) before an array of photodetectors projects the state onto one element of the photon-number basis.
                }
            \label{Experimental setup}
        \end{figure}

        The UBS can be compared to the aforementioned standard boson sampling algorithms. It is proven in SM \cite{sm} how the probability distribution of the UBS reduces to the SBS one when sending $V \rightarrow \mathbb0$, i.e., one obtains the squared permanent of a submatrix of the (now unitary) $U$ block of $\boldsymbol{T}$. The GBS limiting case, i.e., the Hafnian of $\boldsymbol{G}^S$, is more easily recovered by sending $N_{\mathrm{in}}\rightarrow 0$.
        Furthermore, the result in Eq. \eqref{probformulaubs} can also be rewritten using the Fa'a di Bruno's formula as
       \begin{equation}
        \label{superexp}
            \begin{aligned}
                P_{N\rightarrow N}
                ={}&
                \frac{1}{|\det[U]|}
                \sum_{S
                    }
                    \mathrm{Haf}
                    \left(
                        |U^{-1} V |^{\circ 2}_{(\bar{\beta}, \bar{\beta})}
                    \right)
                \\
                {}&\times
                \sum_{
                    \substack{
                            \pi \in \Pi
                            }
                    }
                    \mathrm{Haf}
                    \left(
                        \boldsymbol{G}^S_{[i,j]}
                    \right)
                \prod_{\substack{
                                B \in \pi \\
                                (k,l) \in (i,j)}}
                \left.
                    \frac{
                            \partial^{|B|}
                            \boldsymbol{G}^S_{kl}(\bar x)
                        }
                    {
                        \prod_{r \in B}\partial x_r
                    }
                \right|_{\bar{x} = \bar{0}},
            \end{aligned}
        \end{equation}
        with $\Pi$ being the set of partitions of the indices $\{1,\dots,N-|\bar{\beta}|\}$, $B$ the size of each partition, $i = i_1,\dots,i_{|\pi|} \; , \; j = j_1,\dots,j_{|\pi|} $ indices belonging to each partition $\pi$ and $[i,j]$ standing for the removal of all $i,j$ rows and columns.
        Equation \eqref{superexp} points out how the number of Hafnians that should be computed for each probability grows as the number of partitions of $N-\bar{\beta}$ elements. This quantity scales with the Bell number of $N-\bar{\beta}$, showing a superexponential behavior  in the number of Hafnians and thus an increased complexity.\\
        Our results are generalized to multiple photons per mode by means of MacMahon's theorem \cite{kocharovsky2022hafnian}, by removing/adding multiple rows and columns conditioned to the number of initial as well as the number of measured photons. However, since the complexity of the Hafnian increases with the rank of the matrix \cite{bjorklund2019faster}, any boson sampling implementation usually tries to avoid multiple clicks in the same detector. 
        For realizing UBS, we outline an experimental setup in Fig. \ref{Experimental setup}, in which multiple seeded-generated single photons undergo a Gaussian transformation performed by $K$ independent single-mode optical parametric amplifiers, followed by a Haar-random unitary matrix \cite{meckes2019random} and  a post-selection onto $N$-photon events.
        
    \paragraph{Numerical simulations.---}
    
        Our UBS renders it possible to continuously transition from SBS to GBS, exploring the full intermediate regime. To this end, we simulate multiple rounds of the UBS. We study the $L_1-$distance between the UBS probabilities and the Hafnians, $d_H = |P_{\mathrm{UBS}}-P_{\mathrm{GBS}}|$, and similarly to the permanents, $d_P = |P_{\mathrm{UBS}}-P_{\mathrm{SBS}}|$, both as function of the squeezing in the region $\zeta \in [0, 8]\,\mathrm{dB}$. For derivatives, we harness automatic differentiation techniques widely popular for neural network training tasks, as they are proven to perform with high precision \cite{margossian2019review,griewank2003mathematical}. We compute the distances for 12 values of $\zeta$ in the aforementioned interval, in a range from two to six modes for a saturated scenario, defined as $N = \lceil M/2 \rceil$ and $K = M$. The same was done in a diluted scenario, where $K = N = \lceil{M/2}\rceil$ \cite{sm}.
        
        \begin{figure}
            \centering
              \includegraphics[width=\linewidth]{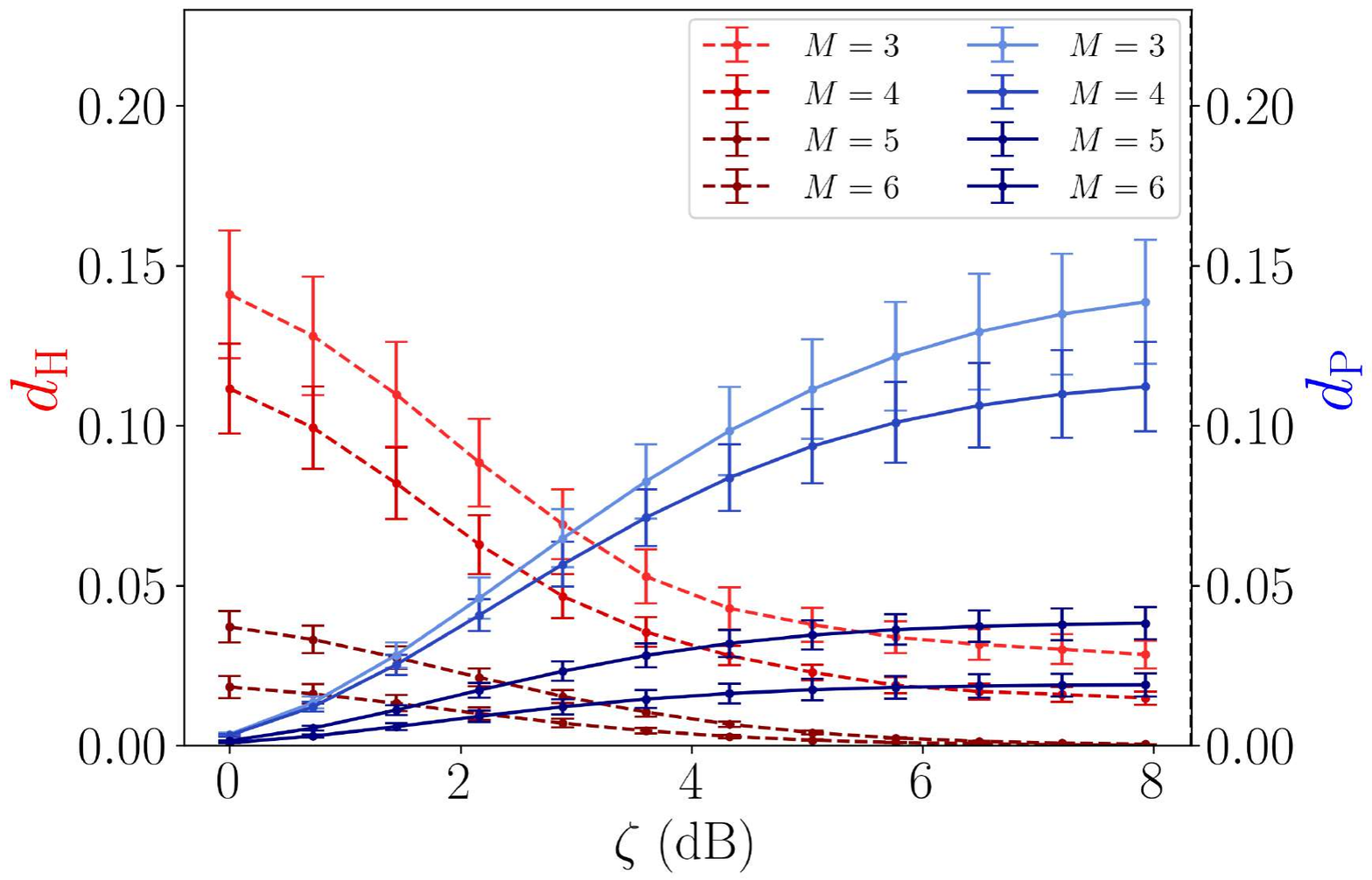}
              \caption{%
                  Average $L_1$-distances between UBS and SBS ($d_{\mathrm{P}}$) and GBS ($d_{\mathrm{H}}$) for different numbers $M$ of modes in the so-called saturated scenario as a function of the squeezing $\zeta$. On the left vertical axis, we display the average distance between UBS and  GBS (dashed, red curves), while on the right one between UBS probabilities and the permanents in SBS (solid, blue curves). We have a minimum distance for $d_{\mathrm{P}}$ and a maximum for $d_{\mathrm{H}}$ at $\zeta = 0\,\mathrm{dB}$ (no squeezing, pure SBS regime), while we have a maximum for $d_{\mathrm{P}}$ and a minimum for $d_{\mathrm{H}}$ at $\zeta = 8\,\mathrm{dB}$ (high value of squeezing intensity, approaching the GBS regime). Statistical fluctuations from the Haar-random selection are displayed via bars.   
              }
            \label{Distances}
        \end{figure}      
        
        We run the simulation multiple times to avoid biases in the choice of particular unitaries or photon patterns, which are randomly generated. The sample variance over Haar random unitaries is given by the average estimator, i.e. 
        $
        \left(
            \sigma/N_{\mathrm{sample}}
        \right)^{1/2}
        $ 
        with $\sigma$ the standard deviation for the average over the runs, and $N_{\mathrm{run}} = 50$. In Fig. \ref{Distances}, we can observe an intersection point between the SBS and the GBS distances around $\zeta \approx 3\,\mathrm{dB}$, representing a midpoint of UBS between GBS and SBS.
        
\begin{figure}
            \centering
            \includegraphics[ width=\linewidth]{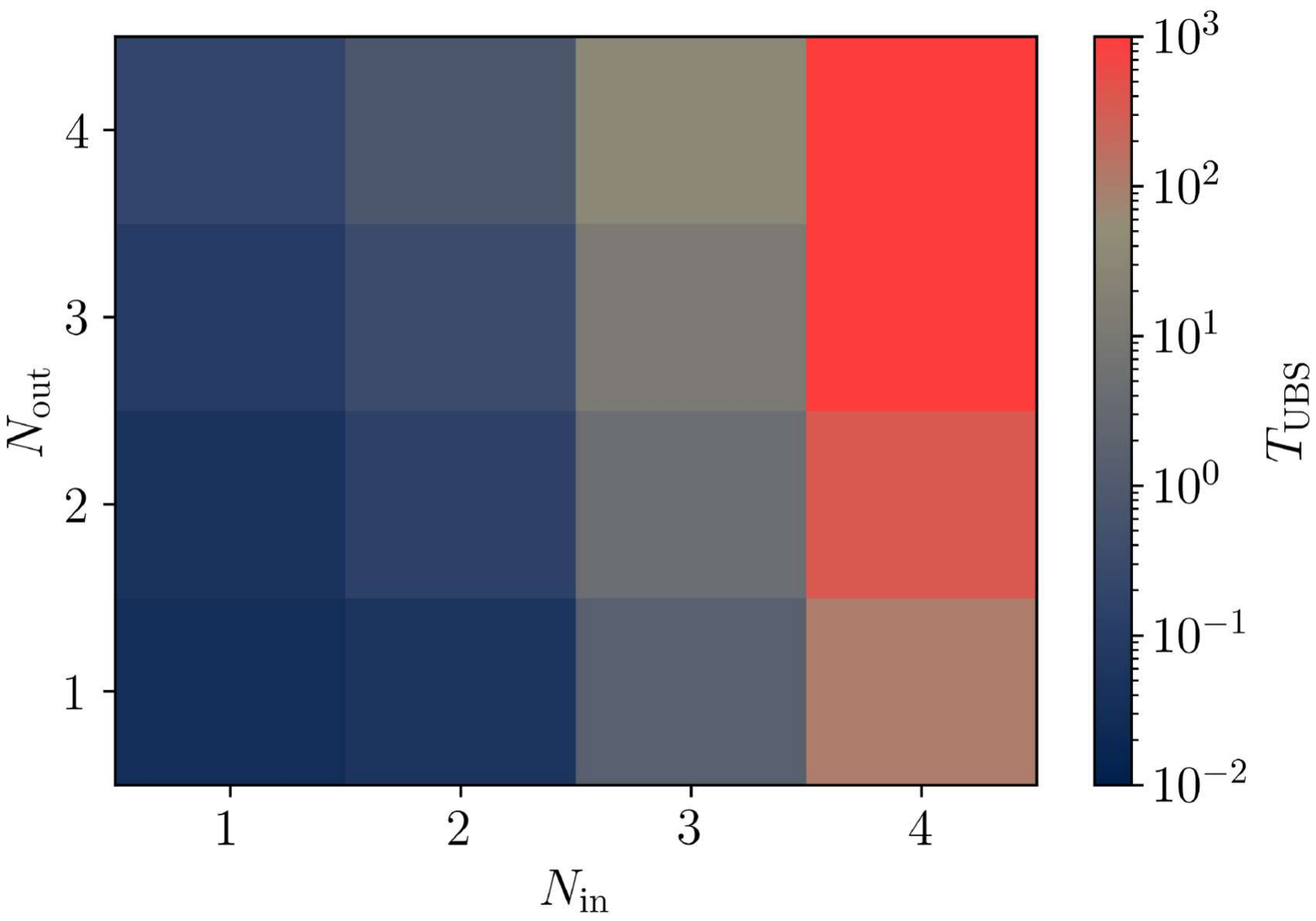}
            \caption{%
            Average computational time for the UBS at $\zeta \approx 3\,\mathrm{dB}$. The average was taken over $10$ rounds of simulation for each $(N_{\mathrm{in}}, N_{\mathrm{out}})$ pairs. The computational time increases more significantly with the input when compared to the output, pointing out the super-exponential behavior, captured through Eq. \eqref{superexp}
            }
            \label{timebenchmark}
        \end{figure}

        To further benchmark the model, we report the computational time as a function of system size in Fig. \ref{timebenchmark}. We fix the squeezing intensity to the midpoint value $\zeta \approx 3\,\mathrm{dB}$, as we observe no changes in the computational time as a function of the squeezing parameter, cf. \cite{sm}. In Fig. \ref{timebenchmark}, we plot the averaged computational time as a function of the input and output photon numbers $N_{\mathrm{in}}$ and $N_{\mathrm{out}}$. We can see that, even for low dimensions, the computational time scales more rapidly with the number of input photons with respect the output photon number, capturing the increased complexity given by the presence of non-Gaussian input states.
        
        Lastly, we study quantum correlations within the UBS. More specifically, we apply the Shchukin-Vogel criterium \cite{shchukin2005inseparability} to a matrix of second-order moments: the presence of negative eigenvalues on a matrix of moments certifies entanglement. We quantify the bipartite entanglement by computing the logarithmic negativity $\log_{2}(2\mathcal{N}+1)$ of the quantum states, with $\mathcal{N}$ being the absolute value of the sum over negative eigenvalues of the partially transposed matrix of moments of the quantum field, i.e. $\mathcal{N} = |\sum_{i} \lambda_{i}^{(<0)}|$.  We focus on a $M = 2d$ modes scenario for $d = 6$ and different number of input photons $N_{\mathrm{in}} = \{2,4,6,8\}$ equally split between the subsystems, a case of particular interest for qudit technologies \cite{bianchi2025predetection}. To this end, we too fix 
        $
        U_{\mathrm{Haar}} 
        = 
        \frac{1}{\sqrt{2}}
        \begin{bmatrix}
             \mathbbm{1} & i\mathbbm{1} \\
             i\mathbbm{1} & \mathbbm{1}
        \end{bmatrix}
        $ where $\mathbbm{1}$ here represents a $d \times d$ identity matrix. The computation of moments is given by the following formula (SM \cite{sm}):
    \begin{equation}
        \label{formula_momenta}
    \begin{aligned}
            {}&
            \langle
                \hat{a}^{k_1}_1\dots \hat{a}^{k_M}_M
                \hat{a}^{\dagger \ l_1}_1\dots\hat{a}^{\dagger \ l_M}_M
            \rangle
            \\
            ={}&
            \left.
                    \frac{(-1)^{\sum_{i=1}^M k_i} \
                        \partial^{\bar{n}}_{\bar{x}} \
                        \partial^{\bar{k}}_{\bar{y}^\ast}\
                        \partial^{\bar{l}}_{\bar{y}}}{\bar{n}!|\det[U]|}
                    \frac{
                        e^{
                            \frac{1}{2}\boldsymbol{y}^{\mathrm{T}}\boldsymbol{G}^{-}\boldsymbol{y}}
                            }
                        {\sqrt{
                            \det{\left[-\boldsymbol{G}\right]}\prod_{i=1}^{M}D_i
                            }
                        }
            \right|_{
                        \bar{y} = \bar{0},
                        \bar{y}^\ast = \bar{0},
                        \bar{x}=\bar{0}
            },
    \end{aligned}
    \end{equation}
        with $\textbf{y}^{\mathrm{T}} = [-y_1^\ast,\dots,-y^\ast_{M},y_1,\dots,y_{M}]$.\\
        \begin{figure}
                \centering
                \includegraphics[width=\linewidth]{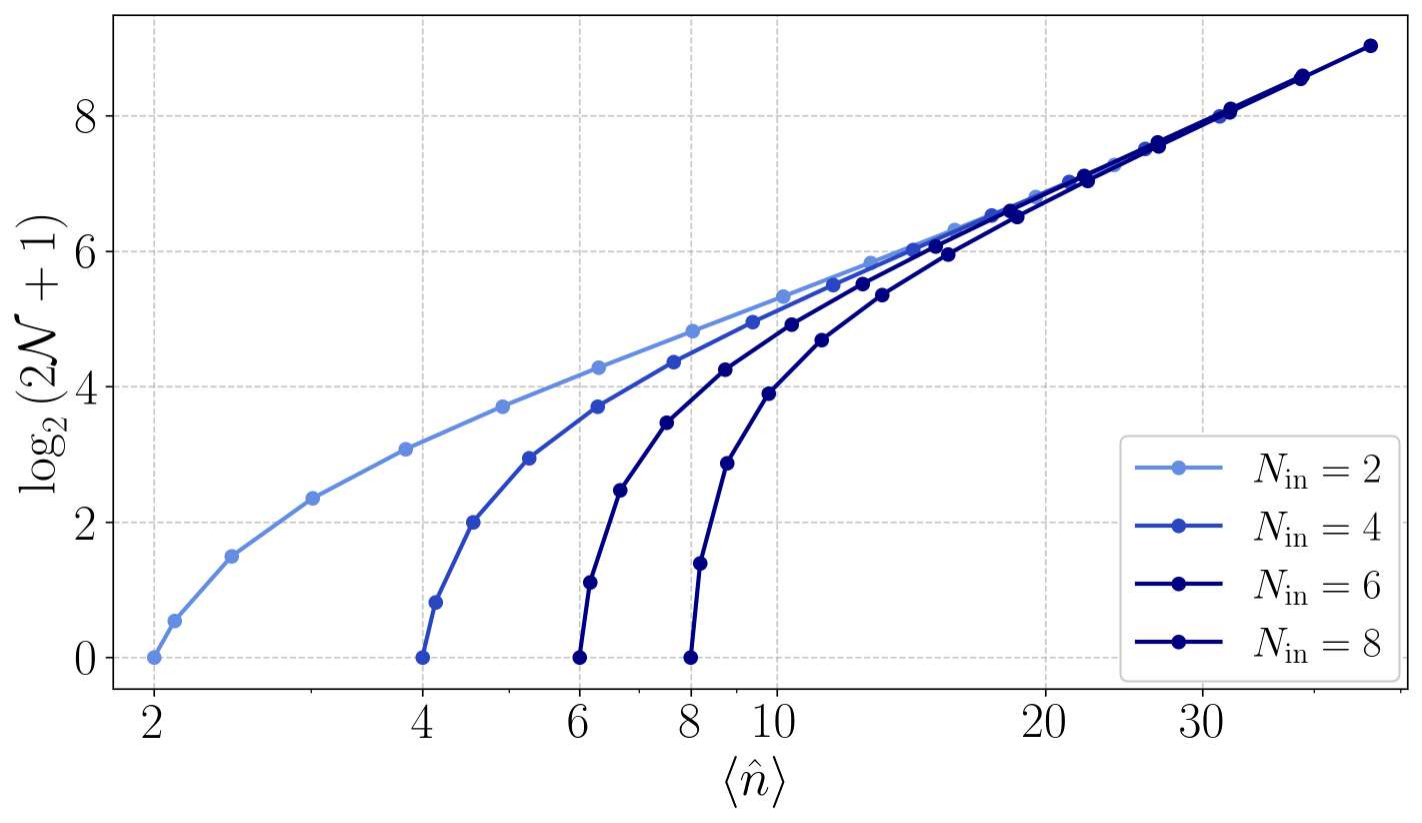}
                \caption{%
                Quantum correlations of the UBS. The logarithmic negativity of the unified boson sampler for the bipartite scenario is plotted as a function of the mean photon number, on a logarithmic scale. Different curves correspond to different total number of input photons in a twelve-mode scenario, while points on each curve represent different squeezing intensities $\zeta$. Deviations on the logarithmic behavior are more marked for low squeezing values and high number of initial photons.
                }
            \label{lognegfigure}
        \end{figure}
        To perform the numerical computations, we rely on a second-order Richardson interpolation \cite{richardson1911ix} of the central finite-difference technique for derivation. (For a comprehensive analysis on the error-estimation of the simulation, see \cite{sm}.)
        In Fig. \ref{lognegfigure}, we show the behavior of the logarithmic negativity as a function of the mean total photon number. Each curve represents a different number of input photons, and each point on the curves stem for a particular squeezing, highlighting the dependency of the entanglement on the two different quantities. From this analysis, we can see how in the high-squeezing regime, the negativity converges to a logarithmic behavior, independently on the number of initial photons. In the low-intensity regime, we observe significant deviations between different scenarios, increasing with $N_{\mathrm{in}}$.\\ 

    \paragraph{Conclusions.---}
    
        In this work, we propose an advanced model of boson sampling, called unified boson sampling, encompassing Gaussian and scattershot samplers as special cases. We develop the formalism to exactly characterize our unified sampler, harnessing the theory of generating functions and representations of the symplectic group. This allows us to compute quantum-physical quantities by means of derivatives of a normally-ordered operator containing the Gaussian nature of the time evolution together with photon-number states. 
        
        In numerical experiments, we quantify the distinctiveness of our UBS and the Hafnian and permanents from common samplers in terms of distances. This is done for different numbers of modes and squeezing values. We further benchmark the complexity in terms of computational time by showing its dependency on input photons, output photons, and squeezing. Finally, we certify the amount of entanglement in the UBS apparatus via the logarithmic negativity and partial transposition, highlighting the different contributions of photon numbers and squeezing. Also, we proposed an experimental setting using state-of-the-art components for realizing the devised protocol.

\section{Acknowledgements}
    The authors are grateful to Nicolás Quesada for useful comments. LB is grateful to Paderborn University for the support.
    LB and DB acknowledge support from the European Union ERC StG, QOMUNE, 101077917. CM acknowledges support from the European Union - NextGeneration EU, ”Integrated infrastructure initiative in Photonic and Quantum Sciences” - I-PHOQS [IR0000016, ID D2B8D520, CUP B53C22001750006].
    LA and JS acknowledge funding through the Ministry of Culture and Science of the State of North Rhine-Westphalia (PhoQC initiative) and the QuantERA project QuCABOoSE.
    
\bibliography{biblio}

\end{document}


\label{SM}

\title{
    Supplemental Material
}

\author{Luca Bianchi}
\affiliation{Department of Physics and Astronomy, University of Florence, 50019, Firenze, Italy}

\author{Carlo Marconi}
\affiliation{Istituto Nazionale di Ottica del Consiglio Nazionale delle Ricerche (CNR-INO), 50125 Firenze, Italy}

\author{Laura Ares}
\affiliation{Theoretical Quantum Science, Institute for Photonic Quantum Systems (PhoQS), Paderborn University, Warburger Stra\ss{}e 100, 33098 Paderborn, Germany}

\author{Davide Bacco}
\affiliation{Department of Physics and Astronomy, University of Florence, 50019, Firenze, Italy}

\author{Jan Sperling}
\affiliation{Theoretical Quantum Science, Institute for Photonic Quantum Systems (PhoQS), Paderborn University, Warburger Stra\ss{}e 100, 33098 Paderborn, Germany}  

\date{\today}

\begin{abstract}
   In this Supplemental Material, we provide detailed insights into the analytical and numerical computations performed to obtain the main results of the work.
   In particular, in Sec. \ref{sec:1A}, we introduce the algebra of Gaussian light, provide the necessary understanding for the Lie algebra rules involved in this work, and derive useful relations that may be useful for analogous tasks.
   In Sec. \ref{sec:1B}, we derive the normal ordering of operators describing multimode, multiphoton states undergoing Gaussian dynamics.
   In Sec. \ref{sec:2}, we introduce the unified boson sampling (UBS) model, showing its consistency with the well-established scattershot and Gaussian boson samplers (cf. Sec \ref{sec:2A}).
   We further prove an identity linking Hafnians to derivatives of determinants (cf. Sec. \ref{sec:2B}).
   In Sec. \ref{sec:2C}, we describe the technique employed for the bipartite entanglement certification.
   Eventually, in Sec. \ref{sec:3}, we discuss details regarding numerical computations.
\end{abstract}
\maketitle

\section{Methods}\label{sec:1}

    \subsection{The algebra of multimode squeezed light}\label{sec:1A}

        Whenever dealing with a semisimple Lie algebra $\mathfrak{a}$ whose generators $T_i$ satisfy commutation relations of the form $[T_i,T_j] = \sum_{k}f_{ij}^kT_k$, it is useful to apply the Cartan decomposition, a process revealing the structure of the algebra and facilitating practical tasks such as Lie algebra classification and representation.
        Specifically, the Lie algebra can be written as $\mathfrak{a} = \mathfrak{h} \bigoplus_{\alpha} \mathfrak{g}_{\alpha}$, where $\mathfrak{h}$ is the Cartan algebra ---the maximal set of mutually commuting generators, denoted as $H_i$--- and $\mathfrak{g}_{\alpha}$ are root subspaces, each associated to a vector $\alpha$ in the root system. 
        For each root vector $\alpha$, the generator $E_{\alpha}$ satisfies
        \begin{equation}
            [H_i, E_{\alpha}] = \alpha(H_i)E_{\alpha} \; \forall H_i \in \mathfrak{h}, 
            \quad
            [E_{\alpha},E_{-\alpha}] = \sum_{i}\alpha(H_i)H_i
            ,\quad\mathrm{and}\quad
            [E_{\alpha},E_{\beta}] = N_{\alpha\beta}E_{\alpha+\beta}
            .
        \end{equation}
         where the first equation also provides the definition of a root generator. Such relations mirror the ladder-operator behavior familiar from quantum mechanics, and the constants $N_{\alpha\beta}$ encode the remaining nontrivial structure of $\mathfrak{a}$. (For further details, see \cite{humphreys2012introduction}.)
         
         In quantum optics, operations such as squeezing, displacement and linear interferometry form an infinite-dimensional representation of $\mathfrak{sp}(2M+2)$ \cite{gilmorecoherent,dictionaryliealgebras}. Through Cartan decomposition, one can identify these continuous-variable operators as finite-dimensional matrices while preserving their algebraic structure. This allows for a simple symbolic handling of multimode quantum states and operators.
        Specifically, one can find a finite-dimensional representation given by a $(M+2)\times(M+2)$ vector space of matrices, and the following mapping gives the relation between the infinite-dimensional space of quantum optical operators and the finite-dimensional representation of $\mathfrak{sp}(2M+2)$ \cite{gilmorecoherent}:
        \begin{equation}
            \begin{split}
            \label{algebrarules}
                &\hat{a}^{\dagger}_i\hat{a}^{\dagger}_j \rightarrow E_{e_i+e_j} \\
                &\hat{a}_i\hat{a}_j \rightarrow E_{-e_i-e_j} \\ 
                &\hat{a}^{\dagger}_i\hat{a}_j\rightarrow E_{e_i-e_j}\\
                &\hat{n}_i +\hat{I}/2 \rightarrow H_i\\
                &\hat{a}^{\dagger}_i \rightarrow E_{e_i-e_{M+1}} \\
                &\hat{a}_i \rightarrow E_{-e_i-e_{M+1}}\\
                &\hat{I} \rightarrow E_{-e_{M + 1} -e_{M + 1}},
                \end{split}
        \end{equation}
        with $e_i \; i = 1,\dots,M$ vectors from the $M$-dimensional canonical basis.
        Among many basis for the finite-dimensional representation of $E$ and $H$, we stick to the one having null entries on the first row and the last column, i.e., the generators can be represented as blocks of the matrix
        \begin{equation}
        \label{matrixrep}
            \begin{pmatrix}
                \begin{array}{c|ccc|ccc|c}
                     0 & 0 & \dots & 0 & 0 & \dots & 0 & 0  \\
                     \hline
                     E_{{e_1-e_{M+1}}}& H_1 & \dots & E_{e_{1}-e_{M}} & E_{e_{1}+e_{M+1}} & \dots & 2E_{e_{1}+e_{1}} &  0\\
                     \vdots & \vdots & \ddots & \vdots & \vdots & \ddots & \vdots & \vdots\\
                     E_{e_{M}-e_{M+1}} & E_{e_{M}-e_1} & \dots & H_{M} & 2E_{e_{M}+e_{M}} & \dots & E_{e_{M}+e_{1}} & 0\\
                     \hline
                     -E_{-e_{1}-e_{M+1}} & E_{-e_{M}-e_{1}} & \dots & -2E_{-e_{M}-e_{M}} & -H_{M} & \dots & -E_{e_{1}-e_{M}} & 0 \\
                     \vdots & \vdots & \ddots & \vdots & \vdots & \ddots & \vdots & \vdots \\
                     -E_{-e_{M}-e_{M+1}} & -2E_{-e_{1}-e_{1}} & \dots & E_{-e_1-e_{M}} & -E_{e_{M}-e_{1}} & \dots & -H_1\\
                     \hline
                     E_{-e_{M+1}-e_{M+1}} & -E_{-e_{M}-e_{M+1}} & \dots & -E_{-e_{1}-e_{M+1}} & -E_{e_{M}-e_{M+1}} & \dots & -E_{e_1-e_{M+1}} & 0 
                \end{array}
            \end{pmatrix}
        \end{equation}
        where, in the above expression, we write the matrices $E, H$ in the positions where their value is $1$, being $0$ elsewhere.

        To give a concrete example, let us consider an expression in terms of the two-modes operators
        \begin{equation}
        \sum_{i,j=1}^2
            \left[
                A_{ij}\hat{a}^{\dagger}_i\hat{a}^{\dagger}_j
                +
                D_{ij}
                    \left(
                        \hat{a}^{\dagger}_i\hat{a}_j + \hat{I}_{ij}/2
                    \right)
                +
                B_{ij}\hat{a}_i\hat{a}_j
                +
                r_i\hat{a}^{\dagger}_i
                +
                l_i\hat{a}_i
                +
                d\hat{I}
            \right].
        \end{equation}
        The terms with $A_{ij}$ and $B_{ij}$ can be seen as quadratic nonlinearities (parametric down-conversion), the terms with $r_i$ and $l_i$ as coherent displacements, the term with $D_{ij}$ as mode transformations (beam splitters and phase shifters), and the terms with $d$ as a constant (global phase shift).
        The mode operators close an $\mathfrak{sp}(6)$ subalgebra, whose finite-dimensional representation may be given in terms of Eq. \eqref{algebrarules}.
        The above expression then reads
        \begin{equation}
            \label{example}
            \begin{pmatrix}
                \begin{array}{c|cc|cc|c}
                     0 & 0 & 0 & 0 & 0 & 0 \\
                     \hline
                     r_1 &  D_{11} & D_{12} & A_{12} & 2A_{11} & 0 \\
                     r_2 & D_{21} & D_{22} & 2A_{22} & A_{21} & 0 \\
                     \hline
                     -l_2 & -B_{21} & -2B_{22} & -D_{22} & -D_{12} & 0 \\
                     -l_1 & -2B_{11} & -B_{12} & -D_{21} & -D_{11} & 0 \\
                     \hline
                     -2d & -l_1 & -l_2 & -r_2 & -r_1 & 0
                \end{array}
            \end{pmatrix}.
        \end{equation}
        Because of canonical commutation relations, matrices describing the mixing of $\hat{a}_{i} \hat{a}_{j}$ and $\hat{a}^{\dagger}_{i} \hat{a}^{\dagger}_{j}$ are symmetric, and thus, their respective blocks in Eq. \eqref{matrixrep} can be written as $2\mathsf{A}$ and $2\mathsf{B}$, where we employ serif and sans serif letters in order to distinguish between the matrices acting on $\hat{a}_i, \,\hat{a}^{\dagger}_j$ and their finite-dimensional representations.
        (Notice how the two representations coincides for matrices describing linear mixing, i.e., $D = \mathsf{D}$.)
        Furthermore, we may introduce the notation $\tilde{\mathsf{A}}$ for a given matrix $\mathsf{A}$, standing for its reflection on the minor diagonal (anti-diagonal).
        The operation carrying on such reflection is unitary, and commutes with transposition, hermitian conjugation, and inversion.
        In such a notation, we can generalize Eq. \eqref{example} to $M$-dimensional objects,
        \begin{equation}
            \begin{pmatrix}
                0 &\vec{0}^\mathrm{T} & \vec{0}^\mathrm{T} & 0 \\
                r & D & 2\mathsf{A} & \vec{0} \\
                -\tilde{l} & -2\tilde{\mathsf{B}} & -\tilde{D} & \vec{0}\\
                0 & -l & -\tilde{r} & 0
            \end{pmatrix},
        \end{equation}
        with $\vec{0}$ a vector of $M$ zeroes.
        Throughout the text, we use $\mathbbm{1}$ for the $M \times M$ identity matrix, $\mathbb{0}$ for the $M \times M$ null matrix, and $\mathrm{T}$, $*$, and ``$-$'' for the transposition, the complex conjugation, and the inverse of a matrix, respectively.

        The convenience of the above representation is apparent when computing products of time-evolution operators that are defined as unitary operators and can be seen as exponentials of generators of $\mathfrak{sp}(2M+2)$.
        Generators corresponding to $\hat{a}_i\hat{a}_j$ and $\hat{a}^{\dagger}_i\hat{a}^{\dagger}_j$ are nilpotent, representations of $\hat{a}^{\dagger}_i\hat{a}_{j}+\hat{I}_{ij}/2$ are block-diagonal and can be easily exponentiated.
        This particularly useful when we perform normal ordering, the common way to express bosonic quantities in quantum optics.
        Recall that normal order sorts creation operators to the left and annihilation operators to the right in operator-valued expressions \cite{vogel2006quantum}.
        To achieve this ordering, it suffices to postulate a normally-ordered expression with unknown coefficients and impose the equality with the starting expression;
        then, comparing the two matrices element-wise yields a complete set of coupled algebraic equations that admit a solution.
        This method lies at the core of the following proposition:
        \begin{proposition}
        \label{prop1}
            Let $A, \, B, \, C,\, D $ be $M \times M$ matrices.
            The following identities hold:
            \begin{subequations}
            \begin{align}
                        \label{eq:Prop1a}
                            e^{\hat{a} A \hat{a}}
                            e^{\hat{a}^{\dagger} C \hat{a}} 
                        ={}&
                            e^{\hat{a}^{\dagger} C \hat{a}}
                            e^{\hat{a} (e^{C})^{\mathrm{T}} A e^C \hat{a}}
                        ,
                        \\\label{eq:Prop1b}
                            e^{\hat{a}^{\dagger}C\hat{a}}
                            e^{\hat{a}^{\dagger}B\hat{a}^{\dagger}}
                        ={}&
                            e^{\hat{a}^{\dagger} e^{C} B (e^{C})^{\mathrm{T}} \hat{a}^{\dagger}}
                            e^{\hat{a}^{\dagger}C\hat{a}}
                        ,
                        \\\label{eq:Prop1c}
                            e^{\hat{a} A \hat{a}} 
                            e^{\hat{a}^{\dagger} B \hat{a}^{\dagger}}
                        ={}&
                            e^{\hat{a}^{\dagger} B D^- \hat{a}^{\dagger}}
                            \prod_{i=1}^{M}
                                \left(
                                    \frac{1}{D_i}
                                \right)^{
                                        \hat{n}_i+\frac{\hat{I}}{2}
                                        }    
                            e^{\hat{a}D^- A \hat{a}}
                        ,
            \end{align}
            \end{subequations}
            with $A,\ B$ symmetric matrices and diagonal entries $D_i=\left(\mathbbm{1}-A^{\mathrm{T}}B^{\mathrm{T}}\right)_{ii}$.
        \end{proposition}

        \begin{proof}
            The first two identities are one the conjugate of the other, so it sufficies to prove the first one.
            Let us postulate the following identity:
                \begin{equation}
                    \label{postulate1}
                            e^{\hat{a} A \hat{a}} 
                            e^{\hat{a}^{\dagger} C \hat{a}} 
                        = 
                            e^{\hat{a}^{\dagger} F \hat{a}}
                            e^{\hat{a} G \hat{a}},
                \end{equation}
            with $F, \, G$ $M \times M$ matrices and $G$ also symmetric.
            The representations of $A, \, C$ in terms of the $\mathfrak{sp}(2M+2)$ subalgebra are
                \begin{equation}
                    \begin{pmatrix}
                        0 & \vec{0}^\mathrm{T} & \vec{0}^\mathrm{T} & 0 \\
                        \vec{0} & \mathbb{0} & \mathbb{0} & \vec{0} \\
                        \vec{0} & -2\tilde{\mathsf{A}} & \mathbb{0} & \vec{0}\\
                        0 & \vec{0}^\mathrm{T} & \vec{0}^\mathrm{T} & 0
                    \end{pmatrix}
                    \quad\mathrm{and}\quad 
                    \begin{pmatrix}
                        0 &\vec{0}^\mathrm{T} & \vec{0}^\mathrm{T} & 0 \\
                        \vec{0} & C & \mathbb{0} & \vec{0} \\
                        \vec{0} & \mathbb{0} & -\tilde{C} & \vec{0}\\
                        0 & \vec{0}^\mathrm{T} & \vec{0}^\mathrm{T} & 0
                    \end{pmatrix}.
                \end{equation}
            Exponentiating those matrices yields
                \begin{equation}
                    e^{\hat{a}A\hat{a}} 
                    \rightarrow 
                            \begin{pmatrix}
                                1 & \vec{0}^\mathrm{T} & \vec{0}^\mathrm{T} & 0 \\
                                \vec{0} & \mathbbm{1} & \mathbb{0} & \vec{0} \\
                                \vec{0} & -2\tilde{\mathsf{A}} & \mathbbm{1} & \vec{0}\\
                                0 & \vec{0}^\mathrm{T} & \vec{0}^\mathrm{T} & 1       
                            \end{pmatrix}
                             \quad\mathrm{and}\quad 
                    e^{\hat{a}^{\dagger}C\hat{a}} \rightarrow 
                            \begin{pmatrix}
                                1 &\vec{0}^\mathrm{T} & \vec{0}^\mathrm{T} & 0 \\
                                \vec{0} & e^{C} & \mathbb{0} & \vec{0} \\
                                \vec{0} & \mathbb{0} & e^{-\tilde{C}} & \vec{0}\\
                                0 & \vec{0}^\mathrm{T} & \vec{0}^\mathrm{T} & 1       
                            \end{pmatrix}.
                \end{equation}
            The same representation holds for $F, \; G$.
            Then, Eq. \eqref{postulate1} reads:
                \begin{equation}
                    \begin{pmatrix}
                                1 & \vec{0}^\mathrm{T} & \vec{0}^\mathrm{T} & 0 \\
                                \vec{0} & e^{C} & \mathbb{0} & \vec{0} \\
                                \vec{0} & -2\tilde{\mathsf{A}}e^{C} & e^{-\tilde{C}} & \vec{0}\\
                                0 & \vec{0}^\mathrm{T} & \vec{0}^\mathrm{T} & 1       
                    \end{pmatrix} = 
                     \begin{pmatrix}
                                1 & \vec{0}^\mathrm{T} & \vec{0}^\mathrm{T} & 0 \\
                                \vec{0} & e^{F} & \mathbb{0} & \vec{0}\\
                                \vec{0} & -2e^{-\tilde{F}}\tilde{\mathsf{G}} & e^{-\tilde{F}} & \vec{0}\\
                                0 & \vec{0}^\mathrm{T} & \vec{0}^\mathrm{T} & 1       
                    \end{pmatrix}.
                \end{equation}
            By equating the blocks of those matrices, we obtain a pair of coupled algebraic equations. From the first of these, it is clear how $ F = C $ and plugging the above  condition into the second yields 
                \begin{equation}
                    \tilde{\mathsf{G}} = e^{\tilde{C}}\tilde{\mathsf{A}}e^{C}.
                \end{equation}
            By taking multiplication, it is now straightforward to check that $e^{\tilde{C}} \tilde{\mathsf{A}}$ coincides with the finite-dimensional $\mathfrak{sp}(2M+2)$ representation of $(e^{C})^{\mathrm{T}}A$.
            Going back to the modes then proves the claim in Eq. \eqref{eq:Prop1a}.
            Hermitian conjugation then yields Eq. \eqref{eq:Prop1b}.

            To prove Eq. \eqref{eq:Prop1c}, we postulate
            \begin{equation}
                \label{postulate2}
                    e^{\hat{a} A \hat{a}} 
                    e^{\hat{a}^{\dagger} B \hat{a}^{\dagger}} 
                =
                    e^{\hat{a}^{\dagger} F \hat{a}^{\dagger}}
                    e^{\hat{a}^{\dagger} E \hat{a}}
                    e^{\hat{a} G \hat{a}},
            \end{equation}
            with $\boldsymbol{E} = \begin{bmatrix}
                \mathrm{diag}(E)/2 & \mathbb{0} \\
                \mathbb{0} & \mathrm{diag}(E)/2 
            \end{bmatrix}$.
            Leveraging again the finite-dimensional representation, Eq. \eqref{postulate2} becomes
            \begin{equation}
                \begin{pmatrix}
                    1 &\vec{0}^\mathrm{T} & \vec{0}^\mathrm{T} & 0 \\
                    \vec{0} & \mathbbm{1} & 2\mathsf{B} & \vec{0} \\
                    \vec{0} & -2\tilde{\mathsf{A}} & \mathbbm{1}-4\tilde{\mathsf{A}}\mathsf{B} & \vec{0}\\
                    0 & \vec{0}^\mathrm{T} & \vec{0}^\mathrm{T} & 1   
                \end{pmatrix} = 
                \begin{pmatrix}
                    1 &\vec{0}^\mathrm{T} & \vec{0}^\mathrm{T} & 0 \\
                                \vec{0} & e^{\mathrm{diag}(E)} - 4\mathsf{F}e^{-\mathrm{diag}(\tilde{E})}\tilde{\mathsf{G}} & 2\mathsf{F}e^{-\mathrm{diag}(\tilde{E})}  & \vec{0} \\
                                \vec{0} & -e^{-\mathrm{diag}(\tilde{E})}2\tilde{\mathsf{G}} & e^{-\mathrm{diag}(\tilde{E})} & \vec{0}\\
                                0 & \vec{0}^\mathrm{T} & \vec{0}^\mathrm{T} & 1
                \end{pmatrix},
            \end{equation}
            from which we see that
            \begin{equation}
                e^{
                    \mathrm{diag}(\tilde{E})
                    } 
                = 
                \left(
                    \mathbbm{1}-
                    4\tilde{\mathsf{A}}\mathsf{B}
                \right)^{-}
                ,\quad
                \tilde{\mathsf{G}} 
                = 
                e^{
                    \mathrm{diag}(\tilde{E})
                    }\tilde{\mathsf{A}}
                ,\quad\mathrm{and}\quad
                \mathsf{F} 
                = 
                \mathsf{B}
                e^{
                    \mathrm{diag}(\tilde{E})
                    }
                .
            \end{equation}
            The first equation readily yields $\mathrm{diag}(E) = -\log(D)$, with $D = \mathrm{diag}\left(\mathbbm{1}-A^{\mathrm{T}}B^{\mathrm{T}}\right)$, while the latter two relations, by matrix inversion, give $\mathsf{F} = \mathsf{B}D^-$ and $\mathsf{G} = D^- \mathsf{A}$.
            Again, going back to the infinite-dimensional representation given by the modes gives $F = B D^- \, , \; G = D^- A$, thus proving the third claim.
        \end{proof}

    \subsection{Normal ordering}\label{sec:1B}

        To achieve a normally-ordered form of the time-evolved generating function, we first of all reorder the unitary operator describing the dynamics.
        For this purpose, we rely on the group theoretical properties of the generators of the dynamical transformation previously discussed.
        Examples of such approach are given for very simple and specific cases, e.g., linear interferometry \cite{disentanglement}, where the Hamiltonians describing the dynamics are mapped through the Jordan-Schwinger theory of angular momentum \cite{schwinger2000quantum} and single-mode as well as two-mode squeezing \cite{caves2020su11}, both represented via finite-dimensional $\mathfrak{su}(1,1)$ subalgebra giving the characteristic hyperbolic nature to squeezing coefficients.

        From a given symplectic matrix $\boldsymbol{T}$, we can always reconstruct its corresponding unitary operator acting on the Hilbert space by introducing an $M\times M$ hermitian Hamiltonian coefficient matrix with the following prescription \cite{quantumofopticalelements}: $H_T=-i\log{(\boldsymbol{T})}\boldsymbol{J}/2$, with $\log{(\boldsymbol{T})}$ denoting the matrix logarithm of $\bm{T}$.
        Then, the unitary operator simply reads
        $\hat{T}=
            \exp{
                \left\{
                    i\boldsymbol{\hat{a}}^{\mathrm{T}}H_{T}\boldsymbol{\hat{a}}
                \right\}
            }
        $.
        Given the Bloch-Messiah decomposition \cite{simonbmdecomposition}, we only need to give expressions for $M$-dimensional unitary matrices $U$, whose Hamiltonian read
        \begin{equation}
            H_U=
                \frac{1}{2}\begin{bmatrix}
                    -i\log{U} & \mathbb{0}\\
                   \mathbb{0} & i\log{U^*}\\
                \end{bmatrix},
        \end{equation}
        and $M$-independent single-mode squeezing transformations, for which we have
        \begin{equation}
            H_{S}= 
                \frac{1}{2}
                    \begin{bmatrix}
                        \mathbb{0}& iZ\\
                        -iZ^*& \mathbb{0}\\ 
                    \end{bmatrix}
                    ,
        \end{equation}
        where 
        $
        Z 
        = 
        \mathrm{diag}(z_1,\dots,z_{M})
        $
        and 
        $
        z_i 
        = 
        \mathrm{arctanh}
            \left(
                \sigma_{v_i}/\sigma_{u_i}
            \right)
        $.
        This allows us to cast the Bloch-Messiah decomposition $\boldsymbol{T} = \boldsymbol{L} \boldsymbol{S} \boldsymbol{R}$ in an operator form
        \begin{equation}
            \label{unitdecompose}
                \begin{split}
                    \hat{T}&=
                        \hat{T}_L\hat{T}_{S}\hat{T}_R \\
                            &=
                            \exp{
                                \left\{
                                    i\sum_{j,k=1}^{M}
                                        H_{L \; j,k}(\hat{a}^{\dagger}_j\hat{a}_k 
                                        +\hat{I}_{jk}/2)
                                \right\}
                                }
                            \exp{
                                \left\{
                                    \sum_{i=1}^{M}
                                        \left(
                                            -\frac{z_i}{2}\hat{a}^{\dagger \; 2}_i
                                            +\frac{z^*_i}{2}\hat{a}^2_i
                                        \right)
                                \right\}
                                }
                            \exp{
                                \left\{
                                    i\sum_{j,k = 1}^{M}
                                        H_{R \; j,k}(\hat{a}^{\dagger}_j\hat{a}_k
                                        +\hat{I}_{jk}/2)
                                \right\}
                                }.
            \end{split}
        \end{equation}
        As previously discussed, the above $M$-modes Gaussian Hamiltonians form a closed subalgebra of $\mathfrak{sp}(2M+2)$, and Eq. \eqref{unitdecompose} can be written in the finite-dimensional representation as
        \begin{equation}
            \begin{split}
            \hat{T} \rightarrow 
                    \exp{
                        \left\{
                            \frac{1}{2}\sum_{j,k=1}^{M}
                                \left(
                                    \log L - \log L^{*\mathrm{T}}
                                \right)_{jk}
                                E_{e_j-e_k}
                        \right\}
                        }
                    \exp{
                        \left\{
                            \sum_{j=1}^{M}
                                \left(
                                    -z_jE_{e_j+e_j}
                                    +z^*_jE_{-e_j-e_j}
                                \right)
                        \right\}
                        }
                    \exp{
                        \left\{
                            \frac{1}{2}
                            \sum_{j,k=1}^{M}
                                \left(
                                    \log R - \log R^{*\mathrm{T}}
                                \right)_{jk}
                                E_{e_j-e_k}
                        \right\}
                        }.
            \end{split}
        \end{equation}
        Then, we can postulate $\hat{T}$ to have the normally-ordered form
        \begin{equation}
        \label{unitaryno}
            \hat{T}_{\mathrm{NO}} \rightarrow 
                    \exp{
                        \left\{
                            \sum_{j,k=1}^{M}
                                B_{jk}E_{e_j+e_k}
                        \right\}
                        }
                    \exp{
                        \left\{
                            \sum_{j=1}^{M}
                                D_{jj}H_j 
                            +\sum_{j\neq k=1}^{M}
                                D_{jk}E_{e_j-e_k}
                        \right\}
                        }
                      \exp{
                        \left\{
                            \sum_{j,k=1}^{M}
                                A_{jk}E_{-e_j-e_k}
                        \right\}
                        }.
        \end{equation}
        We can now compute matrix exponentials, and comparing the two expressions for $\hat{U}$ leads to a system of coupled equations,
        \begin{equation}
            \begin{cases}
                L \sigma_U R = e^{D},\\
                -L
                \mathsf{\sigma_V}
                \tilde{R}^{*T} 
                = 
                2\mathsf{B}
                e^{-\tilde{D}},\\
                \tilde{L}^{*\mathrm{T}}
                \mathsf{\sigma_V}^{*\mathrm{T}}
                R 
                = 
                e^{-\tilde{D}}
                2\tilde{\mathsf{A}},\\
                \tilde{L}^{*\mathrm{T}}
                \tilde{\sigma_U}
                \tilde{R}^{*\mathrm{T}} 
                = 
                e^{-\tilde{D}},
            \end{cases}
        \end{equation}
        where  
        $\mathsf{\sigma_V}=
            \begin{bmatrix}
                0 & \dots & 0 & \sigma_{V_{1}} \\
                0 & \dots & \sigma_{V_2} & 0 \\
                \vdots & \dots & \vdots & \vdots \\
                \sigma_{V_{M}} & 0 & \dots & 0
            \end{bmatrix}$.
        Solving this system results in
        \begin{equation}
            B = -\frac{1}{2}U^{-*\mathrm{T}}V^{\mathrm{T}}
            ,\quad
            A = \frac{1}{2}V^{*\mathrm{T}}U^{-*\mathrm{T}}
            ,\quad\mathrm{and}\quad
            D = \log U^{-*\mathrm{T}}
            .
        \end{equation}
        Interestingly, the symmetry of $B$ and $A$ is guaranteed by the Bloch-Messiah decomposition, i.e., $B = (1/2)L \sigma_{U}^{-1}\sigma_V L^{\mathrm{T}}$ and $A = (1/2)R^{\mathrm{T}}\sigma_V^*\sigma_U^{-*}R$.

       The action of the unitary operator in Eq. \eqref{unitaryno} on the generating function is given by
       \begin{equation}
            \begin{split}
                \hat{T}\hat{E}(\bar{x})\hat{T}^{\dagger}
                =&
                 \exp{
                    \left\{
                        -\frac{1}{2}\hat{a}^{\dagger}U^{-*\mathrm{T}} 
                        V^{\mathrm{T}}
                        \hat{a}^{\dagger}
                    \right\}
                    }
                 \exp{
                    \left\{
                        \log{U^{-*\mathrm{T}}}
                        \left(
                            \hat{a}^{\dagger}\hat{a}
                            +\hat{I}/2
                        \right)
                    \right\}
                    }
                 \exp{
                    \left\{
                        \frac{1}{2}\hat{a}V^{*\mathrm{T}}U^{-*\mathrm{T}}\hat{a}
                    \right\}
                    }\\
                & 
                \times 
                    \hat{E}(\bar{x}) 
                    \exp{
                        \left\{
                            \frac{1}{2}\hat{a}^{\dagger}
                            U^{-}V
                            \hat{a}^{\dagger}
                            \right\}
                        }
                    \exp{
                        \left\{
                            \log{U^{-}}
                                \left(
                                    \hat{a}^{\dagger}\hat{a}
                                    +\hat{I}/2
                                \right)
                            \right\}
                        }
                    \exp{
                        \left\{
                            -\frac{1}{2}\hat{a}
                            V^*U^{-} 
                            \hat{a}
                        \right\}
                        }
                    ,
            \end{split}
        \end{equation}
        where we introduce a multi-index notation, i.e., $\bar{x} = (x_1, \dots, x_{M})$,
        and, by means of the identities derived in Proposition \ref{prop1}, we can move the exponentials until we have all $\hat{a}^{\dagger}_i\hat{a}^{\dagger}_j$-terms on the left and the $\hat{a}_i\hat{a}_j$-terms on the right,
        \begin{equation}
            \label{eqcen1}
            \begin{split}
                \hat{T}\hat{E}(\bar{x})\hat{T}^{\dagger}
                =&
                \left(
                        \prod_{i=1}^{M}
                        \frac{1}{D_i^{1/2}}
                    \right)
                \exp{
                    \left\{
                    -
                    \frac{1}{2}\hat{a}^{\dagger}
                    \left(
                        U^{-*\mathrm{T}}V^{\mathrm{T}}
                        -U^{-*\mathrm{T}}XU^{-}V
                        XD^-U^{-*}
                    \right)
                    \hat{a}^{\dagger}
                    \right\}
                    }\\
                &\times
                    \exp{
                        \left\{
                            \log{U^{-*\mathrm{T}}}
                            \left(
                                \hat{a}^{\dagger}\hat{a}
                                +\hat{I}/2
                            \right)
                        \right\}
                        }
                    \prod_{i=1}^{M}
                        \left(
                            \frac{X_i}{D_i}
                        \right)^{\hat{a}_i^{\dagger}\hat{a}_i}
                    \exp{
                        \left\{
                            \log{U^{-}}
                            \left(
                                \hat{a}^{\dagger}\hat{a}
                                +\hat{I}/2
                            \right)
                        \right\}
                        }
                \\
                &\times
                \exp{
                    \left\{
                    -\frac{1}{2}\hat{a}
                        \left(
                            V^{*}U^{-}
                            -U^{-\mathrm{T}}XD^-V^{*\mathrm{T}}U^{-*\mathrm{T}}XU^{-} 
                        \right)
                        \hat{a}
                    \right\}
                    }
                ,
            \end{split}
        \end{equation} 
        with $X = \mathrm{diag}\left(x_1,\dots,x_M\right)$ and $D_i=\left(1-XU^{-*}V^{*}XV^{\mathrm{T}}U^{-\mathrm{T}}\right)_{ii}$. Finally, we can deal with the term in the third line of Eq. \eqref{eqcen1};
        by applying again the algebra representation, we can rewrite such term as
        \begin{equation}
            \exp{
                \left\{
                    \sum_{ij}
                        \log 
                            \left[
                                U^{-*\mathrm{T}}XD^-U^{-}
                            \right]_{ij}
                            \left(
                                \hat{a}^{\dagger}_i\hat{a}_j
                                +\hat{I}_{ij}/2
                            \right)   
                \right\}
                }
            \exp{
                \left\{
                    -\frac{1}{2}\mathrm{Tr}
                    \left[
                        XD^-
                    \right]
                \right\}
                }.
        \end{equation}
        The matrix $\log \left[ U^{-*\mathrm{T}}XD^-U^{-}\right]$ is hermitian, and thus diagonalizable by a unitary matrix.
        This allows us to write
        \begin{equation}
            \exp{
                \left\{
                    \sum_{ij}
                    \log 
                        \left[
                            U^{-*\mathrm{T}}XD^- U^{-}
                        \right]_{ij}
                    \left(
                        \hat{a}^{\dagger}_i\hat{a}_j
                        +\hat{I}_{ij}/2
                    \right)   
                \right\}
                } = 
            \exp{
                \left\{
                    \frac{1}{2}\mathrm{Tr}
                    \left[ 
                        \log
                            \left(
                            XD^-
                            \right)
                    \right]
                \right\}
            }
            \exp{ 
                \left\{
                    \sum_{ij}
                        \left(
                            U^{-*\mathrm{T}}XD^- U^{-} 
                            -\mathbbm{1}
                        \right)_{ij}\hat{a}^{\dagger}_i\hat{a}_j
                \right\}
                }.
        \end{equation}
        By combining the operator-independent factors, we can recast Eq. \eqref{eqcen1} as
        \begin{equation}
            \label{eqcen2}
            \begin{split}
                \hat{T}\hat{E}(\bar{x})\hat{T}^{\dagger}
                =&\left(
                        \frac{1}{|\det
                            [U]|\prod_{i=1}^{M}D_i^{1/2}}
                    \right)
                \exp{
                    \left\{-
                        \frac{1}{2}\hat{a}^{\dagger}
                            \left(
                                U^{-*\mathrm{T}}V^{\mathrm{T}}
                                -U^{-*\mathrm{T}} X U^{-1}V
                                XD^-U^{-*}
                            \right)
                    \hat{a}^{\dagger}
                    \right\}
                    }\\
                &\;\times
                    :\exp{ 
                        \left\{
                            \hat{a}^{\dagger}
                            \left(
                                U^{-*\mathrm{T}}XD^-U^{-} 
                                -\mathbbm{1}
                            \right)
                            \hat{a}
                        \right\}
                        }:\\
                &\times
                    \exp{
                        \left\{
                            -\frac{1}{2}\hat{a}
                            \left(
                                V^{*}U^{-}
                                -U^{-\mathrm{T}}XD^-V^{*\mathrm{T}}U^{-*\mathrm{T}}XU^{-} 
                            \right)
                            \hat{a}
                        \right\}
                        }
                ,
            \end{split}
        \end{equation}
        with ``$:\cdots:$'' denoting the normal ordering symbol.
        This result encompasses specific examples in multimode linear optics \cite{sperlingdetectorchar}, single-mode squeezing \cite{bianchi2025predetection}, and two-mode squeezing \cite{engelkemeier2020quantum}.

        The time-evolved multimode photon state as used in the main text is therefore given by
            \begin{equation}
                \label{measurement1}
                    \left.
                        \hat{\Pi'}_{\bar{n}}
                        =
                        \frac{\partial^{\bar{n}}_{\bar{x}}}
                            {\bar{n}!}
                        \hat{E}'(\bar{x})
                    \right|_{\bar{x}=\bar{0}},
            \end{equation}
        or, more explicitly, by
            \begin{equation}
            \label{measurement2}
                \left.
                    \hat{\Pi'}_{\bar{n}}
                    =
                    \frac{\partial^{\bar{n}}_{\bar{x}}}{\bar{n}!}
                        :\frac{
                            \exp{ 
                                \left\{
                                    \frac{1}{2}\boldsymbol{\hat{a}}^{\mathrm{T}}
                                    \boldsymbol{G}
                                    \boldsymbol{\hat{a}}
                                \right\}
                                }
                            }
                        {|\det[U]|
                            \sqrt{
                                \prod_{i=1}^{M}D_i}
                        }:
                \right|_{\bar{x}=\bar{0}},
            \end{equation}
        with $\boldsymbol{\hat{a}}^{\mathrm{T}} = [\hat{a}_1, \dots, \hat{a}_{M}, \hat{a}_1^{\dagger},\dots,\hat{a}_{M}^{\dagger}]$, the introduction of
            \begin{equation}
                \boldsymbol{G}=
                \begin{bmatrix}
                    A & B\\
                    B^{\mathrm{T}} & A^*
                \end{bmatrix}
            \end{equation}
        and $A_{ij}^{(*)}$ for the symmetric matrix combining $\hat{a}_{i}^{(\dagger)},\, \hat{a}_{j}^{(\dagger)}$ modes and, likewise, $B_{ij}$ for the hermitian matrix combining $\hat{a}^{\dagger}_i, \, \hat{a}_j$.

\section{Boson sampling}\label{sec:2}

    For the following considerations, it is convenient to describe photodetection through the projection onto the Fock states via the Glauber-Sudarshan representation \cite{mandel1995optical}, harnessing the normal order from the previous section.
    For a specific multimode photon number $\bar m$, the Glauber-Sudarshan representation is given by
    \begin{equation}
        \ket{\bar m}\bra{\bar m}
        =
        \bigotimes_{j=1}^{M}
            \ket{m_j}\bra{m_j}
        =
        \int\prod_{j=1}^{M} 
            d\alpha^{(2)}_j
                \sum_{k_j=0}^{m_j}
                    \frac{1}{k_j!}
                    \binom{m_j}{k_j}
                    \partial^{k_{j}}_{\alpha_{j}}\partial^{k_j}_{\alpha^*_{j}}
                        \left[
                            \delta^{(2)}(\alpha_j)
                        \right]
                    \ket{\alpha_j}\bra{\alpha_j}.
    \end{equation}
    Thus, the probability distribution giving $\bar n$ input photons undergoing Gaussian transformations and photodetection is
    \begin{equation}
        \label{genericprob}
            \mathrm{P}_{\bar n \rightarrow \bar m}
            =
            \mathrm{Tr}[
                \ket{\bar m}\bra{\bar m}
                \hat{\Pi}_{\bar n}
                ]
            =
            \prod_{i,j=1}^{M} 
                \partial^{n_i}_{x_i}
                    \left[
                        \frac{1}{D_i^{1/2}}
                        \sum_{k_j=0}^{m_j}
                            \frac{1}{k_j!}
                            \binom{m_j}{k_j}
                            \partial^{k_{j}}_{\alpha_{j}}\partial^{k_j}_{\alpha_j^*}
                            e^{
                                \boldsymbol{\alpha}^{\mathrm{T}}
                                \boldsymbol{G}
                                \boldsymbol{\alpha}
                            }    
                    \right]_{
                        \substack{
                            \bar{\alpha} = \bar{0}, \bar{\alpha}^* = \bar{0},
                            \bar{x}= \bar{0}}},
    \end{equation}
    where $\boldsymbol{\alpha}^{\mathrm{T}} = [\alpha_1,\dots,\alpha_{M},\alpha_1^*,\dots,\alpha_{M}^*]$.
    For the usual boson sampling scenario, we suppose to have $N$ input photons and $N$ single clicks on photodetectors.
    Firstly, we are going to recover the know regimes of boson sampling.
    It is worth mentioning that any result can be raised to multiphoton input and output states by means of MacMahon's Theorem \cite{kocharovsky2022hafnian}.

        \subsection{Recovering known cases}\label{sec:2A}

             To recover the squared permanent, we need to impose the condition of linear optics, i.e., $V_{ij} = 0$, that further implies $A_{ij}=0$, $|\det
                        [U]| = 1$ and $D_i = 1 \; \forall \, i,j=1,\dots,M $. 
             \begin{equation}
                \mathrm{P}_{\bar n \rightarrow \bar m}
                =
                \prod_{i,j=1}^{M}
                    \partial_{x_i}^{n_i}
                    \left.
                            \partial^{m_{j}}_{\alpha_{j}}
                            \partial^{m_j}_{\alpha_j^*}
                        \exp{
                            \left\{
                                \frac{1}{2}\boldsymbol{\alpha}^{\mathrm{T}}
                                \begin{bmatrix}
                                 \mathbb{0} & B \\
                                B^{\mathrm{T}} & \mathbb{0}\\ 
                                \end{bmatrix}
                                \boldsymbol\alpha
                            \right\}
                        }
                    \right|_{
                        \substack{
                            \bar{\alpha} = \bar{0},
                            \; \bar{\alpha}^* = \bar{0},
                            \bar{x} = \bar{0}}}.
              \end{equation}
            and $B_{ij} = -\sum_{r=1}^{M}U^{-1}_{ir}(1-x_r)U_{rj}^{-*\mathrm{T}}$.
            Applying the Fa'a Di Bruno's formula, we can take the derivatives in $\alpha ,\, \alpha^*$.
            Given the linear mixing from the variables in the exponential, only pairs of derivatives made by one derivative in $\alpha$ and one derivative in $\alpha^*$ yield a contribution to the probability.
            Therefore, we have to take all the second-order derivatives that match the elements from the ensemble of the $N$ $\alpha$s with the $N$ $\alpha^*$s.
            This results in
            \begin{equation}
                \mathrm{P}_{\bar n \rightarrow \bar m}
                =
                \prod_{i=1}^{M}
                \partial_{x_i}^{n_i}
                \left.
                    \left[
                        \sum_{\substack{t=1 
                                    \\\pi_t \in \{1,\dots,N\}}
                                    }^{N!}
                        \prod_{s=1}^{N}
                            \frac{\partial^2}
                                {\prod_{l=1}^{2}
                                    \partial\alpha_l
                                    \partial\alpha^*_l
                                }
                            \left(
                               \alpha^*B\alpha
                            \right)
                    \right]
                \right|_{\substack{
                            \bar{\alpha} = \bar{0},
                            \; \bar{\alpha}^* = \bar{0},
                            \bar{x} = \bar{0}}}.
            \end{equation}
            The derivatives can be reordered by recognizing how the partitions $\pi_t$ are generated by the symmetric group $S_{N}$, and that the surviving terms in the sum are given by the perfect matching pairs $\mathsf{PMP}$ \cite{combinatorialpmp} of derivatives in $\alpha$ and $\alpha^*$.
            In particular, by labelling the derivatives with the corresponding mode, we can reorder the sum by means of a $2N$ dimensional vector $\mu_{j}$ satisfying the conditions $\mu_{j}(2k-1)\leq\mu_{j}(2k)$ and $\mu_{j}(2k-1)\leq\mu_{j}(2k+1)$,
            \begin{equation}
                \mathrm{P}_{\bar n \rightarrow \bar m}
                =
                \left.
                    \prod_{i=1}^{M}
                        \partial_{x_i}^{n_i}
                        \left[ 
                            \sum_{\mu_j\in \mathsf{PMP}}^{N!}
                                \prod_{k = 1}^{N}
                                    B_{\mu_{j}(2k-1), \mu_j(2k)}
                        \right]
                \right|_{\bar{x} = \bar{0}}
            \end{equation}
            The derivatives in $x_i$ then yield the modulus squared of the permanent of the $N\times N$ submatrix $U_S$ given by the input photons and the output click pattern:
            \begin{equation}
                \begin{split}
                    \mathrm{P}_{\bar n \rightarrow \bar m} 
                    &= 
                    \sum_{\mu_j\in \mathsf{PMP}}^{N!}
                        \prod_{i=1}^{M}
                        \partial_{x_i}^{n_i}
                    \left[
                        \prod_{k = 1}^{N}
                            B_{\mu_{j}(2k-1), \mu_j(2k)}
                    \right]_{\bar{x} = \bar 0 }\\
                    &=
                    \sum_{\sigma}
                        \prod_{i=1}^{M}
                            \partial_{x_i}^{n_i}
                    \left[
                        \prod_{p=1}^{N}
                            B_{p, \sigma(p)}
                    \right]_{\bar x = \bar 0}\\
                    &=
                    \sum_{\rho}
                        \sum_{\sigma}
                            \prod_{k=1}^{N}
                            \left[
                                \partial_{x_{\rho(k)}} 
                                B_{k,\sigma(k)}
                            \right]_{\bar{x} = \bar 0}\\
                    &=
                    \sum_{\rho}
                        \sum_{\sigma}
                            \prod_{k=1}^{N}
                                U^{*\mathrm{T}}_{k,\rho(k)}U_{\rho(k),\sigma(k)}\\
                    &=
                    \sum_{\rho}
                        \prod_{k=1}^{N}
                            U^{*\mathrm{T}}_{k,\rho(k)}
                            \sum_{\sigma}
                                \prod_{l=1}^{N}
                                    U_{l,\sigma(l)}\\
                    &=
                    |\mathrm{Perm}[U_S]|^2,
                \end{split}
            \end{equation}
            where we have rewritten the matching pairs in terms of the permutations of $N$ elements in the second line, we made use of Leibnitz rule and the linearity in $x_i$ of the matrix $B$ in the third line, and we relabelled the indices in the product of matrices $U,\, U^{*\mathrm{T}}$ in the last line.

            To recover the Hafnian of the input Gaussian state \cite{hamilton2017gaussian}, it suffices to see that $\alpha_j,\,\alpha_j^*$-derivatives in Eq. \eqref{genericprob} yield only second-order derivatives of the matrix elements. The Fa'a di Bruno's formula then again gives a sum of products of elements of different partitions that turns out to be a sum over perfect matching pairs of derivatives of $\boldsymbol{G}$ elements. Specifically, the derivatives select a submatrix $\boldsymbol{G}^S$ of $\boldsymbol{G}$ by extracting the $j, \, j+M$ rows and columns, and the sum over perfect matching pairs directly gives the Hafnian of $\boldsymbol{G}^S$. In this way, it is straightforward to see that for zero input photons, $\bar n \rightarrow \bar 0$, we take no derivatives, evaluate $\bar x = \bar 0$ and recover the Hafnian of the now purely gaussian state.

    \subsection{Hafnian identity}\label{sec:2B}

        In this subsection, we link the derivatives of a generating function depending on two symmetric matrices to the Hafnian of the Hadamard product of the latter.
        Such an expression is reminiscent of other well-known quantities in the field of boson sampling; see, e.g., Refs. \cite{quesada2018gaussian, bulmer2022threshold}.
        \begin{proposition}
            Let $S,R$ be $M \times M$ symmetric matrices.
            Then, the following formula holds:
                \begin{equation}
                    \partial^{N}_{\beta_{x_1\dots x_{N}}}
                        \left. 
                            \left[
                                \frac{1}{
                                    \sqrt{
                                        \prod_{i=1}^{M}
                                            (\mathbbm{1}-XSXR)_{ii}
                                        }
                                    }
                            \right] 
                        \right|_{\bar{x} = \bar 0} = 
                    \mathrm{Haf}
                    \left[
                        (S\circ R )_{(\bar \beta,\bar \beta)}
                    \right],
                \end{equation}
            where $\beta_{x_1\dots x_{N}}$ is a particular choice of $N$ out of $M$ indices $\bar \beta$ and $(S\circ R )_{(\bar \beta,\bar \beta)}$ denotes the submatrix obtained by extracting $N$ rows and $N$ columns corresponding to the indices in $\bar \beta$.
        \end{proposition}

        \begin{proof}
            Let $N \leq M$ and, without loss of generality, suppose $\beta_{x_1\dots x_{N}} = x_1,\dots, x_{N}$.
            We can formulate the following Gaussian integral identity:
            \begin{equation}
                \frac{1}{
                    \sqrt{
                        \prod_{i=1}^{M}
                            (\mathbbm{1}-XSXR)_{ii}
                        }
                    } = 
                \frac{1}{\pi^{\frac{M}{2}}}
                \int\prod_{j=1}^{M}dy_j
                    e^{
                        -\sum_{i=1}^{M}
                            (\mathbbm{1}-XSXR)_{ii}y_i^2
                        } \; .
            \end{equation}
            With this identity, we can write
            \begin{equation}
                \begin{split}
                    \partial_{x_1\dots x_{N}}
                        \left[ 
                            \frac{1}{\pi^{\frac{M}{2}}}
                            \int\prod_{j=1}^{M}dy_j
                                e^{
                                    -\sum_{i=1}^{M}
                                        (\mathbbm{1}-XSXR)_{ii}y_i^2
                                    }
                        \right]_{\bar x=\bar 0} 
                        &= 
                        \left.
                            \frac{1}{\pi^{\frac{M}{2}}}
                            \int\prod_{j=1}^{M}dy_j
                                \partial_{x_1\dots x_{N}}
                                \left[ 
                                    e^{
                                        -\sum_{i=1}^{M}(\mathbbm{1}-XSXR)_{ii}y_i^2
                                        }
                                \right]
                        \right|_{\bar{x} = \bar 0}\\
                        &=
                        \left.
                            \frac{1}{\pi^{\frac{M}{2}}}
                            \int\prod_{j=1}^{M}dy_j
                                \sum_{\substack{
                                        \pi \in \Pi \\ 
                                        \{1,\dots,N\}}
                                        }
                                    \prod_{B \in \pi}
                                        \frac{\partial^{|B|}}{\prod_{l \in B}\partial x_l}
                                        \left[ 
                                            -\sum_{i=1}^{M}
                                                (\mathbbm{1}-XSXR)_{ii}
                                        \right]
                        \right|_{\bar x=\bar 0} \\
                        &= 
                        \left.
                            \frac{1}{2}
                            \sum_{i=1}^{M}
                                \sum_{\substack{
                                    \pi \in \Pi \\ 
                                    \{1,\dots,N\}}
                                    }
                                    \prod_{B \in \pi}
                                        \frac{\partial^{|B|}}
                                        {\prod_{l \in B}
                                            \partial x_l
                                        }
                                        \left[
                                            -(\mathbbm{1}-XSXR)_{ii}
                                        \right]
                        \right|_{\bar x= \bar 0},
                \end{split}
            \end{equation}
            where, in the first line, given the convergence ensured by $0\leq x_i \leq 1 \: \forall \; i \in 1,\dots M $, we swap the $x$ derivatives and integration, in the second line, we apply the Fa'a di Bruno's formula, and, in the last line, we employ the well-known Gaussian formula $\int dy y^2 e^{-y^2} = \pi/2$.

            We can write $(\mathbbm{1}-XSXR)_{ii} = \sum_{m = 1}^{M}\left(\delta_{im} - S_{im}R_{im}x_ix_m\right)$ and recognize that $i \neq m$ are the only nontrivial terms, given the nature of first derivatives.
            Furthermore, the symmetry of $S, \; R$, combined with the sums over $i, \; m$ compensate the $1/2$ with double counting.
            Furthermore, given the quadratic nature of the term inside the derivatives and the evaluation of $\bar x = \bar 0$, the size of each partition is $|B| = 2 \; \forall \; \pi \in \Pi$.
            This also imposes $N$ to be even, and the number of elements in each partitions to be $N/2$.
            It turns out that there are $(N-1)!!$ such partitions,
            \begin{equation}
                \sum_{j=1}^{(N-1)!!}
                    \prod_{k=1}^{N/2} 
                        (S \circ R)_{l_1 l_2},
            \end{equation}
            with $l_1, \, l_2$ belonging to one of the subset $k$ of the partition $j$.
            In particular, the partition is selected by the indices of the fixed partition, extracting an submatrix $N \times N$ submatrix selected by the indices of the derivatives, $\bar \beta$.
            Now, the sum over the possible partitions can be regarded as a sum over perfect matching pair of indices belonging to finite sets of $\{1,\dots, N/2\}$.
            We can reorder the product over the indices in each partition in a vector $\mu_j(k)$ and finally write
            \begin{equation}
                \begin{split}
                    \partial^{N}_{\beta_{x_1\dots x_{N}}}
                    \left.
                        \left[
                            \frac{1}{
                                \sqrt{
                                    \prod_{i=1}^{M}
                                        (\mathbbm{1}-XSXR)_{ii}
                                        }
                                    }
                        \right]
                    \right|_{\bar{x} = \bar 0}
                    &= 
                        \sum_{\mu_j \in \mathsf{PMP}}^{(N-1)!!}
                            \prod_{k=1}^{N/2} 
                                [(S \circ R)_{(\bar \beta,\bar \beta)}]_{\mu_j(2k-1), \mu_j(2k)} \\
                    &= 
                    \mathrm{Haf}
                        \left[
                            (S \circ R)_{(\bar \beta,\bar \beta)}
                        \right],
                \end{split}
            \end{equation}
            thus completing the proof.
        \end{proof}

    \subsection{Bipartite entanglement}\label{sec:2C}

        To apply the Shchukin-Vogel criterium for multipartite continuous-variables quantum systems \cite{shchukin2005inseparability}, we need to build a matrix of moments that becomes negative under partial transposition when entanglement is detected.
        In the following, we focus on bipartite systems, with a partitioning  $P_1 \cap P_2 =  \emptyset $, $P_1 \cup P_2 = M$, $M=2d$ and modes $1,\dots,d \in P_1$, $d+1,\dots,2d \in P_2$ so that only $d$ modes are involved in the part that undergoes transposition.
        We again employ the generating function formalism to organize such moments.
        For any expectation value of the type $
        \langle
                        \hat{a}^{k_1}_1\dots \hat{a}^{k_M}_M
                        \hat{a}^{\dagger \ l_1}_1\dots\hat{a}^{\dagger \ l_M}_M
        \rangle
        $, we may write
        \begin{equation}
            \left.
                    \langle
                        \hat{a}^{k_1}_1\dots \hat{a}^{k_M}_M
                        \hat{a}^{\dagger \ l_1}_1\dots\hat{a}^{\dagger \ l_M}_M
                    \rangle
                = 
                \partial^{\bar{n}}_{\bar{x}}
                \left[
                    \frac{(-1)^{\sum_{i=1}^M k_i}
                        \, \partial^{\bar{k}}_{\bar{y}^*} 
                        \, \partial^{\bar{l}}_{\bar{y}}
                        }
                    {\bar{n}!
                    |\det[U]|
                    \sqrt{\prod_{i=1}^{M}D_i}
                    }
                    \langle
                        e^{-y^{*\mathrm{T}}\hat{a}}
                        e^{y^{\mathrm{T}}\hat{a}^{\dagger}}
                        :e^{\frac{1}{2}
                            \boldsymbol{\hat{a}}^{\mathrm{T}}
                            \boldsymbol{G}
                            \boldsymbol{\hat{a}}
                        }:
                    \rangle
                \right]
            \right|_{\substack{
                        \bar{y} = \bar{0},\bar{y}^* = \bar{0},
                        \bar{x}=\bar{0}}}.
        \end{equation}
        Focusing on the bracket, we can see how it is readily in a normally-ordered form.
        Thus, we can compute the scalar product in the coherent basis.
        By applying a multivariate Gaussian integration after the change of basis
        \begin{equation}
            \begin{bmatrix}
                \mathcal{R}(\alpha)\\
                \mathcal{I}(\alpha)
            \end{bmatrix} = 
            \begin{bmatrix}
                \mathbbm{1} & i \mathbbm{1} \\
                \mathbbm{1} & -i \mathbbm{1}
            \end{bmatrix} 
            \begin{bmatrix}
                \alpha\\
                \alpha^*
            \end{bmatrix},
        \end{equation}
        with $\mathcal{R}(\alpha), \mathcal{I}(\alpha)$ the real and immaginary parts of $\alpha$, we then end up with the following generating function for the moments:
        \begin{equation}
            \label{formula_momenta}
                   \left.
                        \langle
                            \hat{a}^{k_1}_1\dots \hat{a}^{k:M}_M
                            \hat{a}^{\dagger \ l_1}_1\dots\hat{a}^{\dagger \ l_M}_M
                        \rangle 
                    =
                    \frac{(-1)^{\sum_{i=1}^M k_i} \ 
                                    \partial^{\bar{n}}_{\bar{x}} \ 
                                    \partial^{\bar{k}}_{\bar{y}^*}\
                                    \partial^{\bar{l}}_{\bar{y}}
                        }
                    {\bar{n}!
                    |\det{\left[U\right]}|
                    }
                    \frac{
                        e^{
                            \frac{1}{2}\boldsymbol{y}^{\mathrm{T}}
                            \boldsymbol{G}^{-}
                            \boldsymbol{y}
                            }
                        }
                    {\sqrt{
                        \det{\left[-\boldsymbol{G}\right]}
                        \prod_{i=1}^{M}D_i}
                    }
                    \right|_{\substack{
                                        \bar{y} = \bar{0},\bar{y}^* = \bar{0},\bar{x}=\bar{0}}}.
        \end{equation}
        From the above expression, by a symmetry argument, we can see how moments with an odd number of ladder operators vanish, outlining a precise structure for the matrix of partially transposed moments.
        The Shchukin-Vogel method prescribes a matrix of moments in a graded-antilexicographical order, and shows how partial transposition consists on swapping between the creation and annihilation operators of the $P_1$ party that got transposed.
        If we restrict ourselves to fourth order moments, the resulting matrix is a $[(2d+1)^2-2d + 1]/2 \; \times \; [(2d+1)^2-2d + 1]/2$ dimensional matrix, with a block structure like
        \begin{equation}
            \boldsymbol{M} =\begin{bmatrix}
                1 & 0_{2d} & s \\
                0_{2d}^{\mathrm{T}} & S & 0_{2d \times d(2d-1)} \\
                s^{*\mathrm{T}} & 0_{d(2d-1) \times 2d} & F
            \end{bmatrix}
        \end{equation}
        with
        $s$ a $d(2d-1)$ array made by $d(d-1)/2$ moments of the type $\langle\hat{a}^{\dagger}_{i}\hat{a}^{\dagger}_{j}\rangle \; i,j \in P_1$ , $d^2$
        moments of the type $\langle \hat{a}^{\dagger}_{i}\hat{a}_{j}\rangle \; i \in P_1, \; j \in P_2$, and $d(d-1)/2$ 
        moments of the type $\langle \hat{a}_{i}\hat{a}_{j}\rangle \; i,j \in P_2$.
        $S$ is a $2d \times 2d$ matrix of second order moments and $F$ a $d(2d-1) \times d(2d-1)$ matrix of fourth order moments. The moments belonging to the above submatrices can be obtained by intersecting the ones belonging to the first row and the first column, e.g. $S_{11} = \langle \hat{a}_1\hat{a}^\dagger_1\rangle$ and $F_{11} = \langle \hat{a}_1\hat{a}_2\hat{a}_1^\dagger\hat{a}^{\dagger}_2 \rangle$.
        To prove the negativity, we compute the eigenvalues and consider the modulus of the sum of the negative ones $\lambda_i^{(<0)}$ for the sake of the quantification of the thereby verified entanglement.

\section{Simulations}\label{sec:3}

    \subsection{Boson sampling simulation}\label{subsec:3A}

\begin{figure}
    \centering
    \includegraphics[width=0.5\linewidth]{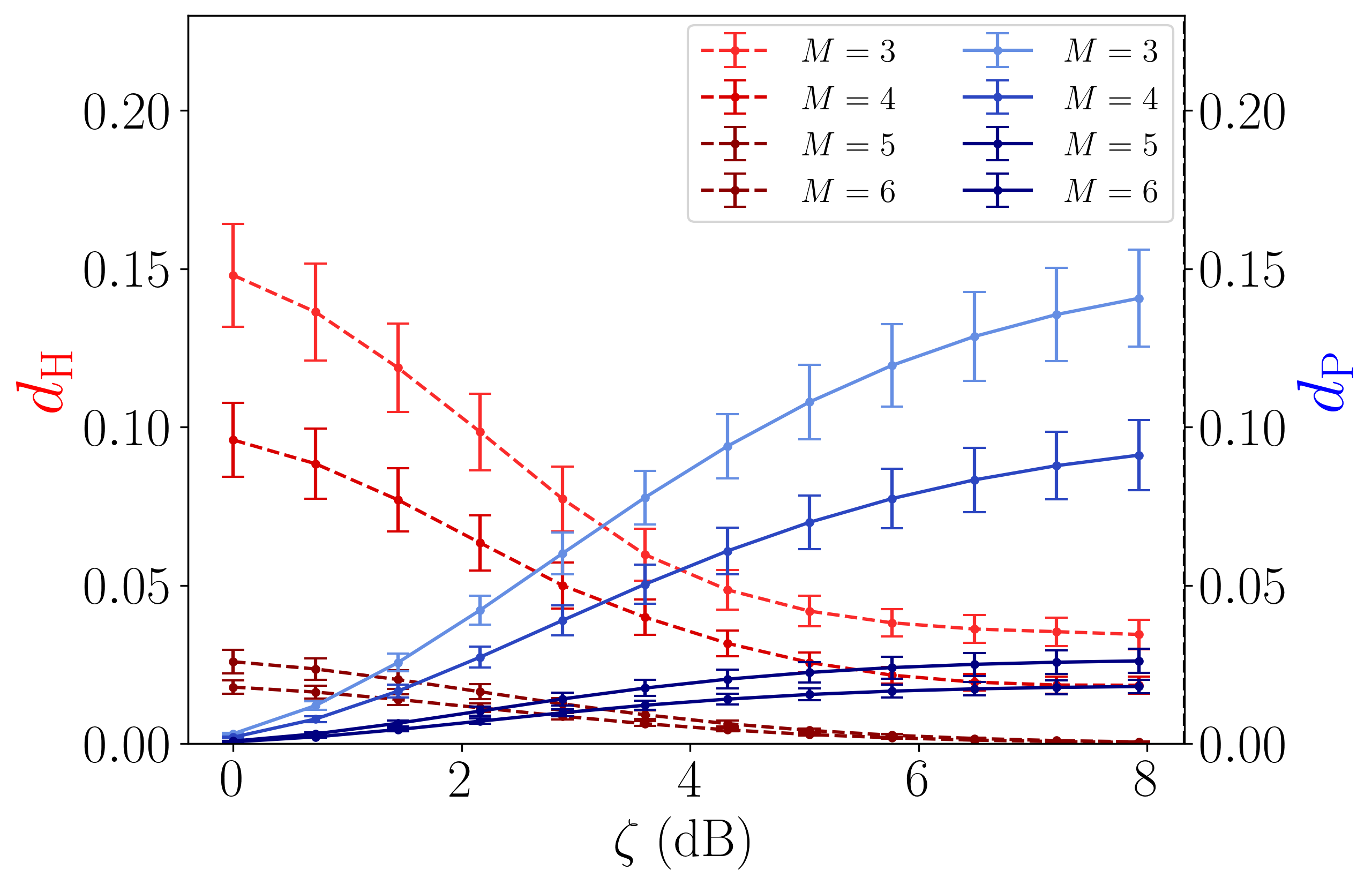}
    \caption{%
        Average $L_1$-distances between probability outcomes computed with scattershot, Gaussian and UBS for different modes in the dilute scenario.
        As in the saturate regime, we plot the average distance between UBS and SBS (solid, blue curves) on the left, while on the right we show the average distance between UBS and GBS (dashed, red curves).
        By comparing the above plot with the one presented in the main text, we can see how the distances do not depend significantly on the number of squeezed modes.
    }\label{dilute}
\end{figure}

        We perform simulations for UBS probabilities in two regimes.
        The first one, that we call saturated, has each mode squeezed by an optical parametric amplifier $K = M$ and $N = \lceil M/2 \rceil$.
        The second one, which we refer to as dilute, has the number of photons and the number of squeezers set as $K = N = \lceil{M/2}\rceil$, and the average $L_1$ distances from the scattershot boson sampling (SBS) and Gaussian boson sampling (GBS) protocols is shown in Fig. \ref{dilute}. Comparing Fig. \ref{dilute} with Fig. 2 of the main text, one notices no difference between the distances computed in the two scenarios, highlighting the leading importance of the system size in the scaling of the distances $d_{\mathrm{P}}$ and $d_{\mathrm{H}}$.  

\begin{figure}
    \centering
    \includegraphics[width=\linewidth]{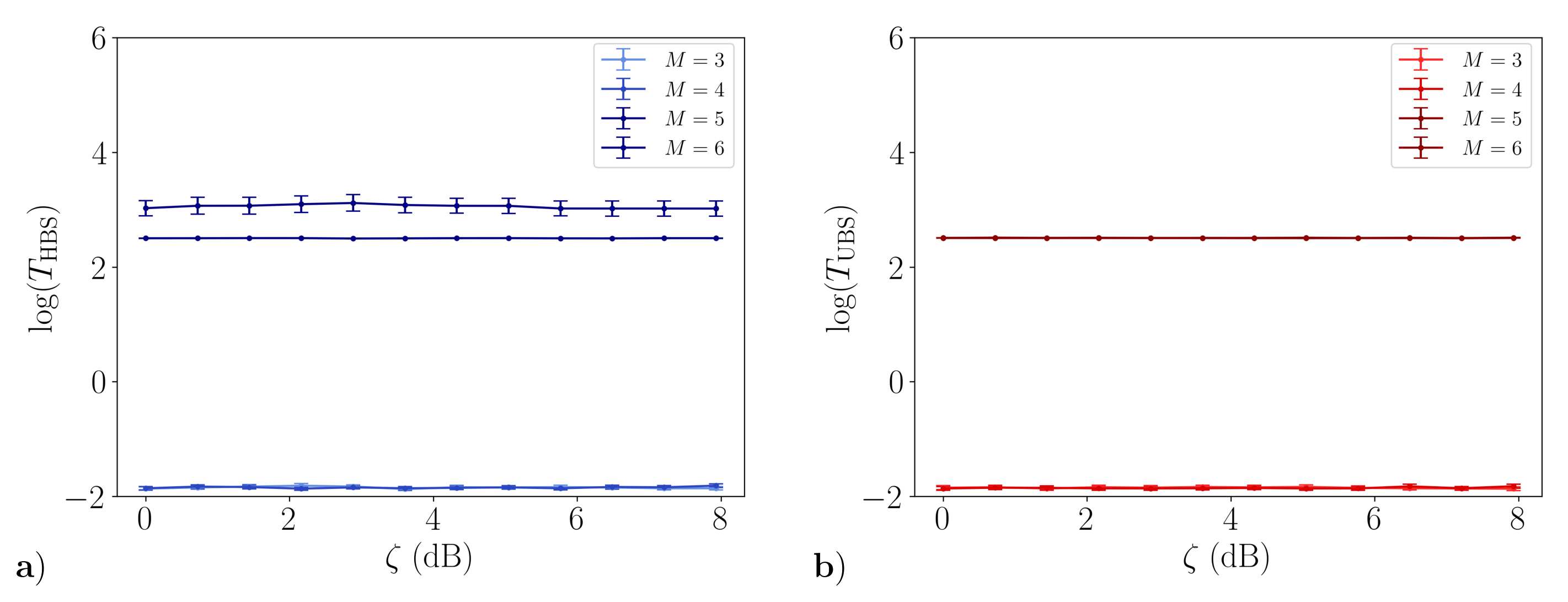}
    \caption{%
        Average logarithmic computational time of UBS in a) the diluted scenario and b) the saturated.
        We observe the dependency on the system size, hinting the increasing complexity with the number of involved photons, but no dependency on the squeezing parameter.
    }\label{time}
\end{figure}

        Analogous conclusions can be drawn by plotting the computational time as a function of the squeezing intensity. In Fig. \ref{time} we observe how the times computed for both dilute and saturated scenarios are squeezing-independent. The above plots justify the complexity benchmark in terms of computational time given in Fig. 3 of the main text, where one can see the scaling of the same quantity in terms of the system size at a fixed squeezing value. 

    \subsection{Entanglement certification}\label{sec:3B}

        To obtain derivatives, we rely on a second-order Richardson extrapolation applied to the central method for finite differences.
        We pick $h = f^{1/(2+n)}h_{\mathrm{scale}}$, with $f \simeq 10^{-16}$ the float precision, $n$ the number of derivatives taken, and $h_{\mathrm{scale}} = 1$ for $2, \, 4$ derivatives and $h_{\mathrm{scale}} = 0.9$ for $ 6, \, 8$ derivatives.
        We give an estimate of the error propagating to the logarithmic negativity coming from numerical approximations and truncation errors.

        Firstly, we benchmark our code by computing mean photon numbers, for which analytical expressions are known.
        We notice how the absolute error scales in the following way with the number of derivatives: $\epsilon_{\mathrm{abs}}^{\mathrm{avg} \; (2)} \simeq 10^{-8}$, $\epsilon_{\mathrm{abs}}^{\mathrm{avg} \; (4)} \simeq 10^{-6}$, $\epsilon_{\mathrm{abs}}^{\mathrm{avg} \; (6)} \simeq 10^{-5}$, $\epsilon_{\mathrm{abs}}^{\mathrm{avg} \; (8)} \simeq 10^{-4}$.
        We overestimate the error by assigning the above errors uniformly to the elements of the matrix $\boldsymbol{M}$ although we know that almost half of its elements are zero, and we do not compute them numerically.
        By the Weyl inequality, we can then upper bound the uncertainty on the eigenvalues $\lambda$ of $\boldsymbol{M}$ by the 2-norm of the matrix $\boldsymbol{E}^{(n)}$ whose elements are  $\epsilon_{\mathrm{abs}}^{\mathrm{avg} \; (n)}$.
        We can pessimistically estimate that $\epsilon_{\lambda}^{(n)} \leq ||\boldsymbol{E}^{(n)}||_{2} \leq \epsilon_{\mathrm{abs}}^{\mathrm{avg} \; (n)}[(2d+1)^2-2d + 1]/2] \simeq \epsilon_{\mathrm{abs}}^{\mathrm{avg} \; (n)} 10^{2}$.
        By applying the propagation of uncertainties, we can estimate the uncertainty over the mean photon number $\epsilon_{\langle\hat{n}\rangle} \simeq \epsilon_{\mathrm{abs}}^{\mathrm{avg} \; (n)} 2d$ and over the logarithmic negativity $\log_2(2\mathcal{N}+1)$, i.e., $\epsilon_{\log_2{(2\mathcal{N}+1)}} = \epsilon_{\lambda}^{(n)}\sqrt{n_{\lambda^{(< 0)}}}/|\sum_{i}\lambda_i^{(< 0)}|$, with $n_{\lambda^{(< 0)}}$ the number of negative eigenvalues.

\bibliography{biblio}